\documentclass[journal]{IEEEtran}
\usepackage{cite}

\usepackage[pdftex]{graphicx}

\usepackage{amsmath}
\usepackage{mathtools}

\usepackage{array}

\usepackage[caption=false,font=footnotesize]{subfig}

\begin{document}
\title{Reduction in Circulating Current with Improved Secondary Side Modulation in Isolated \\Current-Fed Half Bridge AC-DC Converter}

\author{Manish Kumar,~\IEEEmembership{Student Member,~IEEE,}
        Sumit Pramanick,~\IEEEmembership{Member,~IEEE,}\\
        and B K Panigrahi,~\IEEEmembership{Senior Member,~IEEE}}
\maketitle

\begin{abstract}
Current-fed half bridge converter with bidirectional switches on ac side and a full bridge converter on dc side of a high frequency transformer is an optimal topology for single stage galvanically isolated ac-dc converter for onboard vehicle charging application. AC side switches are actively commutated to achieve zero current switching (ZCS) using single phase shift modulation (SPSM) and discontinuous current phase shift modulation (DCPSM). Furthermore, zero voltage turn-on (ZVS) is achieved for dc side switches. Compared to SPSM, DCPSM maintains a constant peak current in the converter throughout the grid cycle of ac mains voltage. However, constant peak current contributes to a high circulating current near the zero crossings of ac mains voltage and also at light load conditions. This paper proposes an improved discontinuous current phase shift modulation (IDCPSM) to increase the efficiency of the converter across different loading conditions. A dual control variable is adopted to actively reduce the circulating current  while maintaining soft switching of both ac and dc side switches across the grid cycle of ac mains voltage. A 1.5 kW laboratory prototype has been developed to experimentally validate the analysis, design and improvement in performance for different loading conditions.          
\end{abstract}
\begin{IEEEkeywords}
Current-fed half bridge, ac-dc converter, single stage, improved discontinuous current phase shift modulation (IDCPSM)
\end{IEEEkeywords}

\section{Introduction}
\IEEEPARstart{E}{lectric} vehicles (EVs) have gained tremendous popularity against their internal combustion engine (ICE) counterparts over the past decade. As EVs are becoming more common, ac-dc power converters are being explored for onboard charging applications (OBCs). OBCs provides the user with the convenience of using an ac utility outlet to charge the EVs battery. The design of an OBC is mainly driven by electrical and volumetric efficiency \cite{khalighGlobalTrendsHighPower2019}. These ac-dc power converters should meet power quality standards on the grid side as per IEEE 519 \cite{IEEERecommendedPractice2014}. Such converters should also have galvanic isolation between the ac side (grid) and the dc side (vehicle battery) as per Underwriters Laboratories (UL) 2202 safety standard \cite{UL2202Standard}. 

The most general approach to attain the above requirements is a two-stage solution. An active power factor correction circuit as the first stage followed by an isolated dc-dc converter as second stage\cite{xueDualActiveBridgeBased2015, kwonElectrolyticCapacitorlessBidirectional2017}. The isolated DC-DC converter is realized by a dual active bridge (DAB) in \cite{xueDualActiveBridgeBased2015} and series resonant converter in \cite{kwonElectrolyticCapacitorlessBidirectional2017}. However, the two-stage approach increases device count and losses, contributing to lower efficiency, along with increase in control complexity and volume. To overcome the disadvantages of the two-stage solution, several single-stage approaches, which could be either voltage-fed or current-fed, have been reported in the literature\cite{evertsOptimalZVSModulation2014, weiseSingleStageDualActiveBridgeBasedSoft2014, jauchCombinedPhaseShiftFrequency2016, tausifSinglestageIsolatedElectrolytic2019}. 

Current-fed isolated converters have some distinct advantages over the voltage-fed isolated converters, such as lower input current ripple, inherent short circuit protection, wide soft-switching operating range and ease of current control \cite{gnanasambandamCurrentFedMultilevelConverters2017, wuBidirectionalBuckBoost2020}. Work presented in \cite{panOverviewComprehensiveComparative2020} exhaustively compared the various isolated current-fed (ICF) topologies in terms of electrical and volumetric efficiency and reported that the L-L type  isolated current-fed half-bridge (ICFHB) is an optimal topology among the different ICF topologies. Thus, an L-L type single stage single phase ICFHB ac-dc converter using bidirectional switches proposed in \cite{prasannaNovelBidirectionalSinglephase2017} is the focus of this paper. This converter shown in Fig. \ref{fig1} is conceptually derived from an L-L type ICFHB dc-dc converter \cite{rathoreAnalysisDesignExperimental2013}. 

However, one major drawback of such converters is that a huge voltage spike appears at the switching node caused by an abrupt change in the current through the high frequency transformer (HFT) winding. It occurs at the instant when the boost inductor current finds a path through the HFT windings. A common technique to mitigate the problem of voltage spike is to use an additional active clamp circuit. An active clamp circuit consisting of two active switches and a clamping capacitor was proposed in \cite{rathoreAnalysisDesignExperimental2012} for L-L type ICFHB. This clamp circuit assisted in zero voltage switching (ZVS) of all the switches. Another active clamp circuit consisting of a single switch, two diodes, a clamping capacitor and a resonating inductor was proposed in \cite{moraesActiveClampedZeroCurrentSwitching2020} that ensures zero current switching (ZCS) of the main primary side switches. This increased component count in the auxiliary clamp circuit attributes to higher losses. 

Another technique is the modulation based approach also known as secondary side modulation (SSM) \cite{rathoreAnalysisDesignExperimental2013, prasannaNovelZeroCurrentSwitchingCurrentFed2013, panImprovedModulationScheme2020, balImprovedModulationStrategy2018, zhangDecoupledDualPWMControl2019}. The secondary side voltage-fed full bridge (VFFB) is used for active current commutation at the primary side current-fed half bridge (CFHB). This is achieved by using the reflected output voltage to completely transfer the boost inductor current from the commutating leg to the non-commuting leg through the HFT. The reflected output voltage is applied across the primary side of the HFT for a sufficiently long duration to force the current through the commutating leg in the reverse direction, through the body diode. When the body diode current naturally reaches zero, the switch is turned-off achieving ZCS. Also, body diode conduction in the secondary side ensures ZVS turn-on of secondary side switches. The merit of this technique is that it does not require any additional circuits.

A single phase shift modulation (SPSM) for an L-L type ICFHB is proposed in \cite{rathoreAnalysisDesignExperimental2013}. The phase shift $\phi$ is between the primary and secondary side bridge gating signals. Soft switching is achieved in both bridge switches for a wide operating range. However, when implemented in the case of the converter shown in Fig. \ref{fig1} with ac input, the circulating current flowing through the switches and the HFT is substantially high near the zero crossing of the ac mains voltage and at light load conditions, attributing to poor efficiency of the converter.

A discontinuous current single phase shift modulation (DCPSM) is proposed in \cite{prasannaNovelZeroCurrentSwitchingCurrentFed2013} where the current through the HFT drops to zero and a  resonance occurs between the output capacitance of the dc side switches and the leakage inductance of the transformer. The circulating current flowing through the devices is still relatively large near the zero crossing of the ac mains voltage. A discontinuous current dual phase shift modulation (DCDPSM) for an L-L type ICFHB dc-dc converter is proposed in \cite{panImprovedModulationScheme2020}, where unsymmetrical switching pattern is applied to dc side switches to avoid resonance. However, a five control variable were adopted to reduce the circulating current value at light load conditions making the control architecture complex.  

This paper proposes an improved discontinuous current single phase shift modulation (IDCPSM) with a simpler unsymmetrical switching pattern. The objective is to minimize the circulating current, flowing through the switches and the HFT across the grid cycle of ac mains voltage and also at different loading conditions. This is achieved by modulating the duty of the dc side VFFB in every switching interval. The duty of ac side CFHB is modulated to maintain unity power factor (UPF). This  dual control variable approach keeps the control architecture simple and easy to implement. ZCS turn-off for ac side switches and ZVS turn-on for dc side switches is maintained across the grid cycle. Reduced RMS current through ac and dc side switches as well as through the HFT is achieved contributing to lower conduction losses with significant improvement in the efficiency of the converter.

This paper is organized as follows. Section II provides a detailed description of the schematic of ICFHB ac-dc converter. Operational stages with the proposed IDCPSM are described and compared with the conventional SSM in Section III. Section IV shows the experimental results of the implemented IDCPSM on the developed laboratory prototype and the comparative performance evaluation among the different SSM.  Lastly, Section V concludes this paper.

\section{ICFHB AC-DC Converter Description}
\begin{figure}[!tbp]
  \centering
    \includegraphics[width=\linewidth]{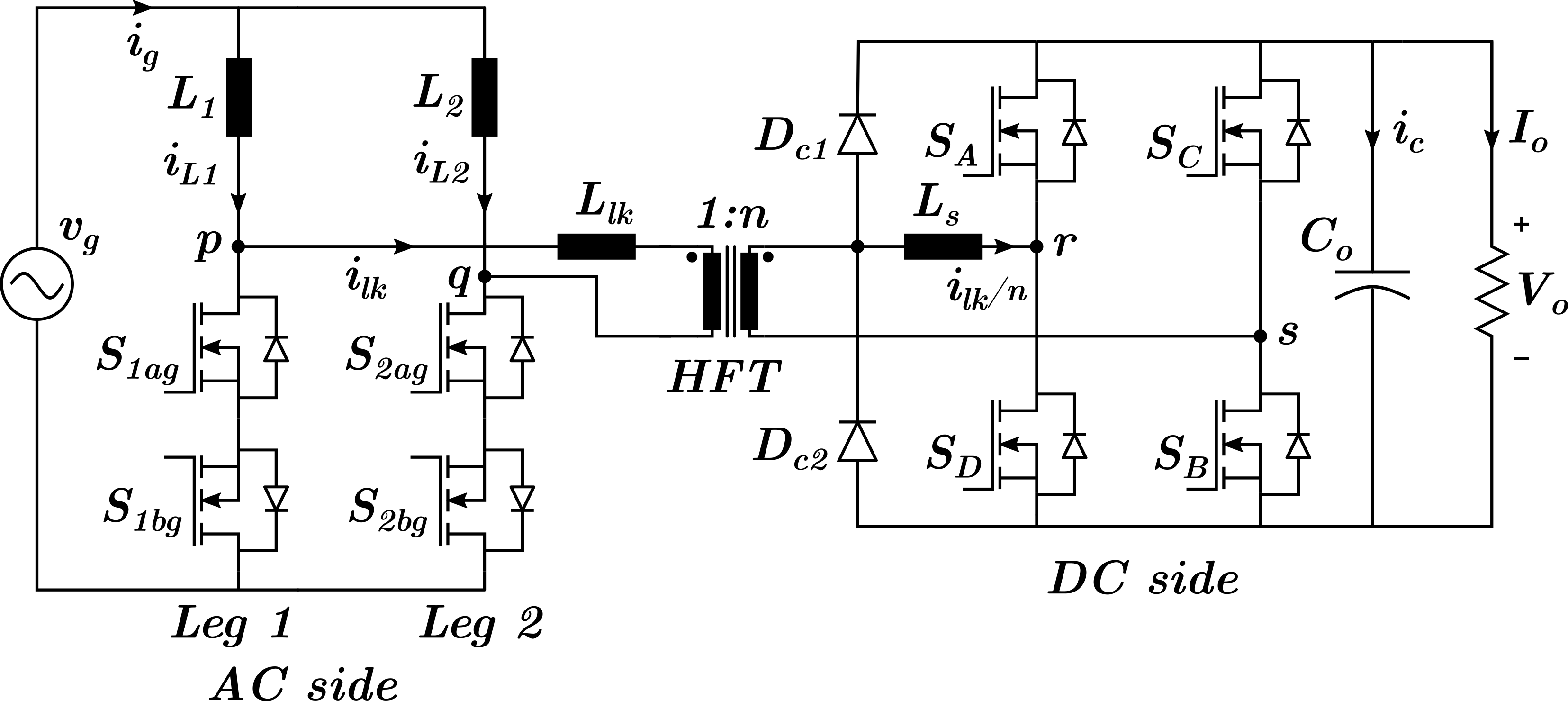}
    \caption{L-L type current fed half bridge single stage ac-dc converter.}
        \label{fig1}
\end{figure}

The schematic of a single phase single stage galvanically isolated current-fed dual active bridge ac-dc converter is shown in Fig. \ref{fig1}. It consists of a current-fed half bridge converter with bidirectional switches on the ac side and a full bridge converter on the dc side, connected through a high frequency transformer (HFT).  $L_{1}$ and $L_{2}$ are the two boost inductors. $L_{lk}$ is the leakage inductance of the HFT. An external series inductor $L_{s}$ is connected on the dc side to achieve the required total series inductance $L_{t} = L_{lk}+L_{s}/n^2$. $C_o$ is the output filter capacitor.     
\par 
The ac side switches are annotated as $S_{xyg}$, where $x$ denotes leg number and $y=a$ or $b$. $S_{xag}$ are switched at high frequency in the positive half cycle of grid voltage and remains continuously on in the negative half cycle. While $S_{xbg}$ are switched at high frequency in the negative half cycle of grid voltage and remains continuously on in the positive half cycle. The gate signals between ac side leg $1$ and leg $2$ switches are phase-shifted by $180^{\circ}$. $S_{xyg}$ is switched at a duty ratio $d_1$ always greater than $0.5$, ensuring  conduction overlap between the legs. Battery side switches are annotated as $S_p$, where $p=A, B, C$ or $D$. $S_A$ and $S_B$ are switched at a fixed duty of 0.5. $S_C$ and $S_D$ are switched at a duty ratio of $d_2$. $S_p$ turn-off is synchronised with the turn-off of $S_{xyg}$ in the following way. In the positive half cycle of the grid voltage, $S_{C,D}$ and $S_{A,B}$ turn off is synchronised with $S_{1ag}$ and $S_{2ag}$ respectively. In the negative half cycle, $S_{C,D}$ and $S_{A,B}$ turn off is synchronised with $S_{2bg}$ and $S_{1bg}$ respectively. During the turn-off duration of ac side switches, a resonance occurs between its output device capacitance and the total series inductance, $L_{t}$. In order to suppress this parasitic ringing, HFT is designed to have very minimum leakage inductance $L_{lk}$. Diode clamps $D_{c1}$ and $D_{c2}$ are connected between the HFT terminals and external series inductor at dc side, thereby clamping the ac side switch voltage to $V_o/n$. These diode clamps incur lower losses than the damping resistor used in \cite{prasannaNovelBidirectionalSinglephase2017}.

\begin{figure*}
     \centering
     \subfloat[\label{fig:90}]
	{
         \centering
         \includegraphics[width=0.31\linewidth]{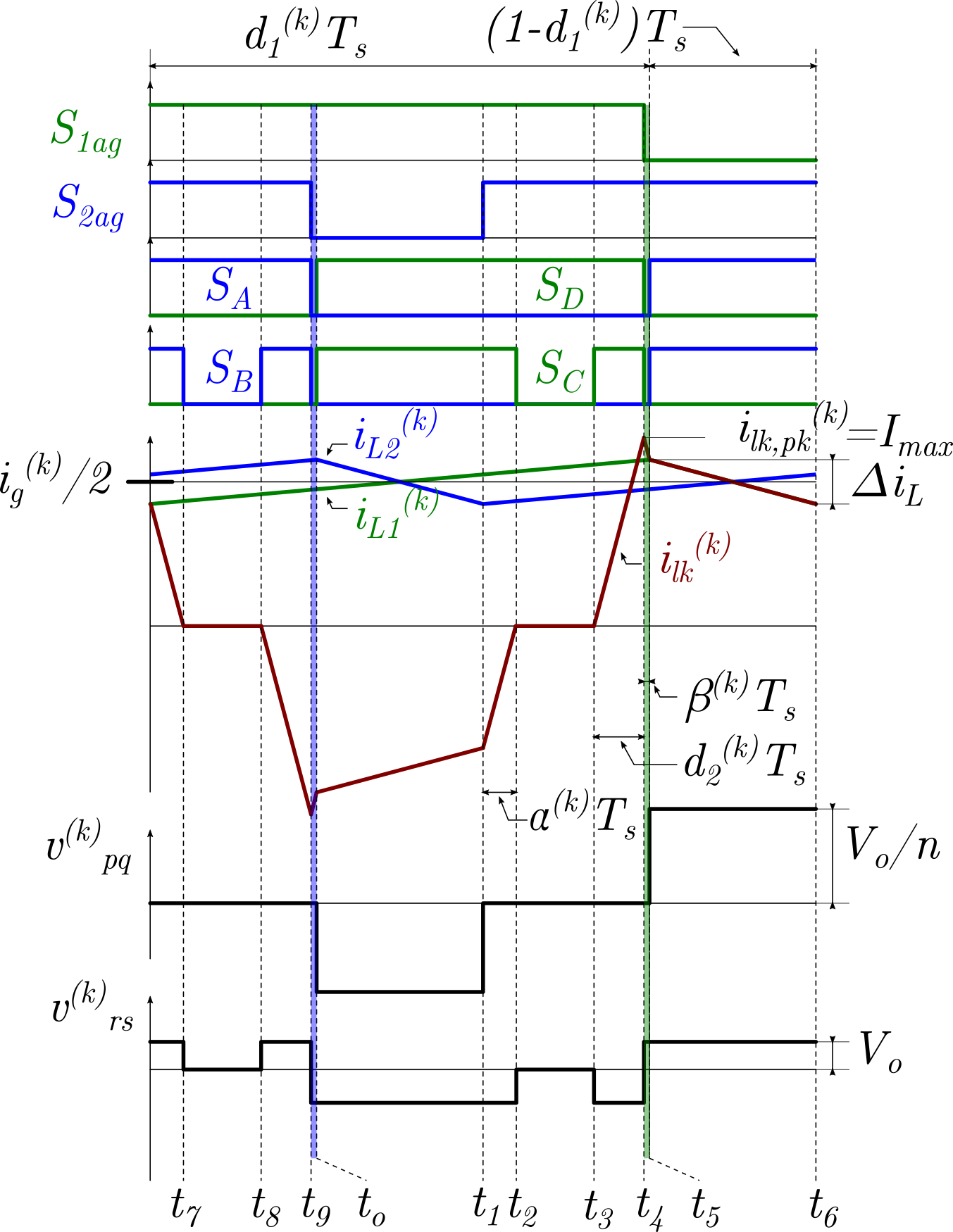}
	}     \hfill
     \subfloat[\label{fig:10}]
	{
         \centering
         \includegraphics[width=0.31\linewidth]{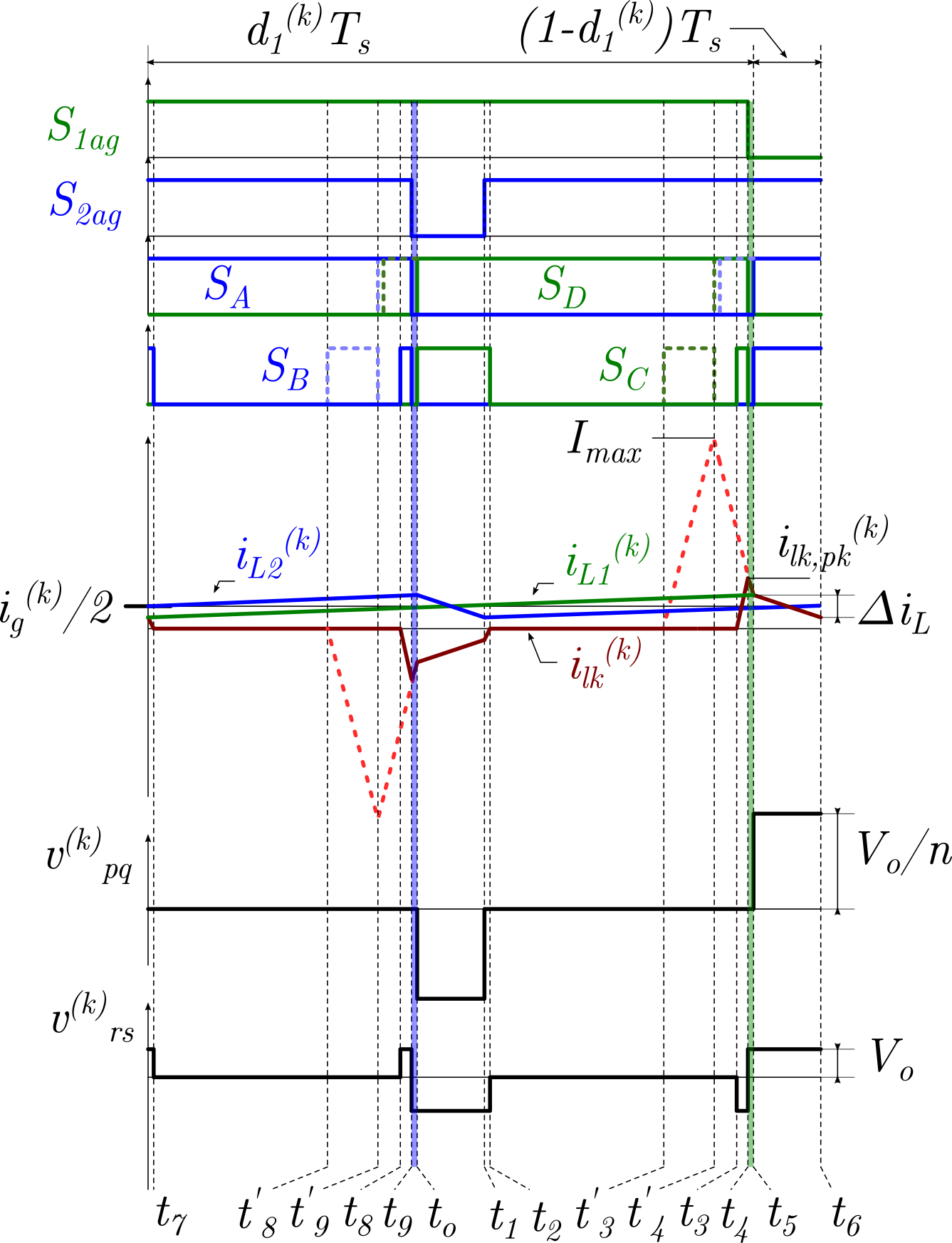}
	}     \hfill
     \subfloat[\label{fig:10_current}]
	{
         \centering
         \includegraphics[width=0.31\linewidth]{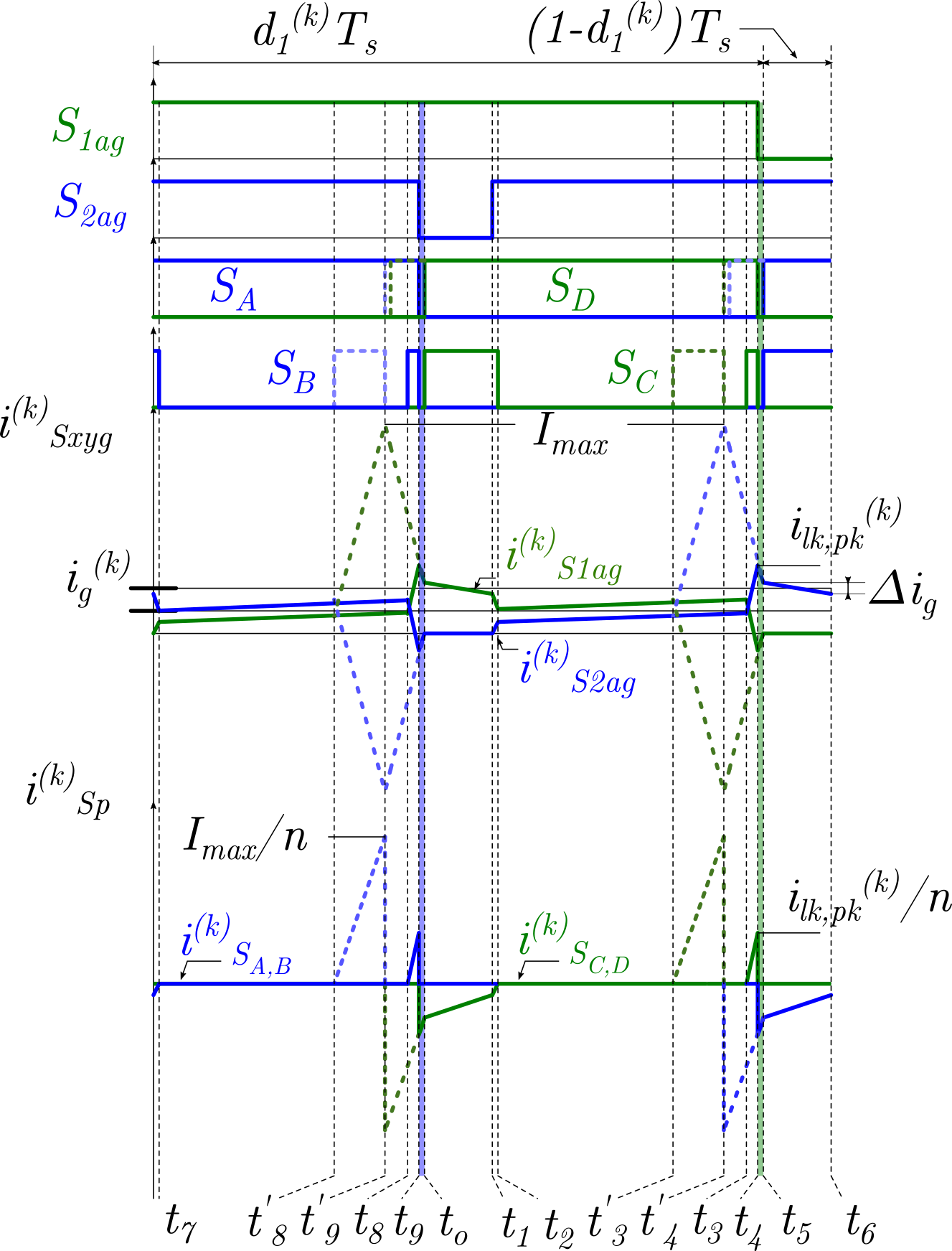}
	}
        \caption{Operating waveform of ICFHB ac-dc converter in a switching interval and in the positive half cycle of ac mains voltage (a) at $\omega \tau=90^{\circ}$ peak of ac mains voltage, (b) at $\omega \tau=10^{\circ}$ near zero crossing of ac mains voltage, (c) ac side switch current $i_{Sxyg}$ and dc side switch current $i_{Sp}$ at $\omega \tau=10^{\circ}$ near zero crossing of ac mains voltage}
\end{figure*}

\section{Operation stages with IDCPSM}
To simplify analysis, it has been assumed that the converter is operating at unity power factor (UPF). Further, input ac mains voltage $v_g^{(k)}(\tau)$ is assumed constant in a switching interval $k$. Variable with superscript $k$ is changing every next switching interval and variable without the superscript $k$ is assumed constant throughout the converter operation.   
\begin{align}
v_g^{(k)}(\tau) = V_{m}sin(\omega \tau)\label{vg}\\
i_g^{(k)}(\tau) = \frac{2P_o}{V_m}sin(\omega \tau)\label{ig}
\end{align}
AC side duty cycle is given by
\begin{align}
d_{1}^{(k)} = \frac{V_o-n|v_g^{(k)}(t)|}{V_o}\label{duty1}
\end{align}

where, $\tau = k*T_s$ and $T_s$ is the time period of a switching cycle, $\omega = 2\pi/t_g$, $t_g$ is the time period of a grid cycle, $P_o$ is the output power and $V_o$ is the output voltage. Fig. \ref{fig:90} and Fig. \ref{fig:10}, shows the operating waveform of the converter at the peak and near the zero crossing in the positive half cycle of ac mains voltage respectively. Fig. \ref{fig:10_current}, shows the  ac side switch current $i_{Sxyg}$ and dc side switch current $i_{Sp}$ near the zero crossing in the positive half cycle of ac mains voltage. 
Inductor current $i_{L1}^{(k)}(t)$ is given by Eq. \ref{inductor}. 
\begin{equation}
\begin{split}
&i_{L1}^{(k)}(t)=\\
&\left\lbrace \begin{matrix*}[l] \frac{1}{2}(i_g^{(k)}-\frac{v_g^{(k)}}{L_1}d_1^{(k)}T_s)+\\\frac{v_g^{(k)}}{L_1}[t-(k-1)T_s]& (k-1)T_s < t \leq (k-1+d_1^{(k)})T_s\\\\ 
\frac{1}{2}(i_g^{(k)}+\frac{v_g^{(k)}}{L_1}d_1^{(k)}T_s)+\\\frac{v_g^{(k)}}{L_1}[t-(k-1+d_1^{(k)})T_s]\\-\frac{V_o}{nL_1}[t-(k-1+d_1^{(k)})T_s]&(k-1+d_1^{(k)})T_s < t < kT_s\\\\ \frac{1}{2}(i_g^{(k+1)}+\frac{v_g^{(k+1)}}{L_1}d_1^{(k+1)}T_s)& t = kT_s \end{matrix*} \right.
\end{split}
\label{inductor}
\end{equation}

\begin{figure*}[!ht]
     \centering
     \subfloat [\label{fig:t0-t1}]{
         \centering
         \includegraphics[width=0.32\linewidth]{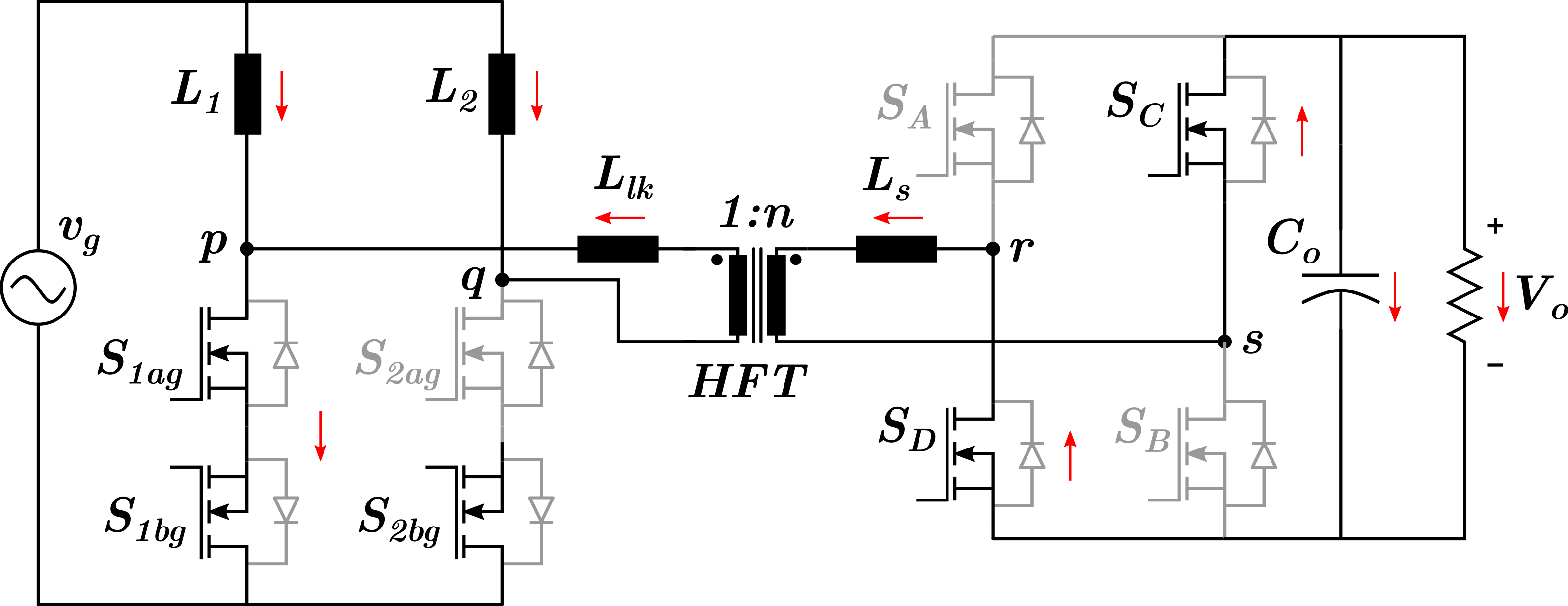}}
     \hfill
     \subfloat [\label{fig:t1-t2}]{
         \centering
         \includegraphics[width=0.32\linewidth]{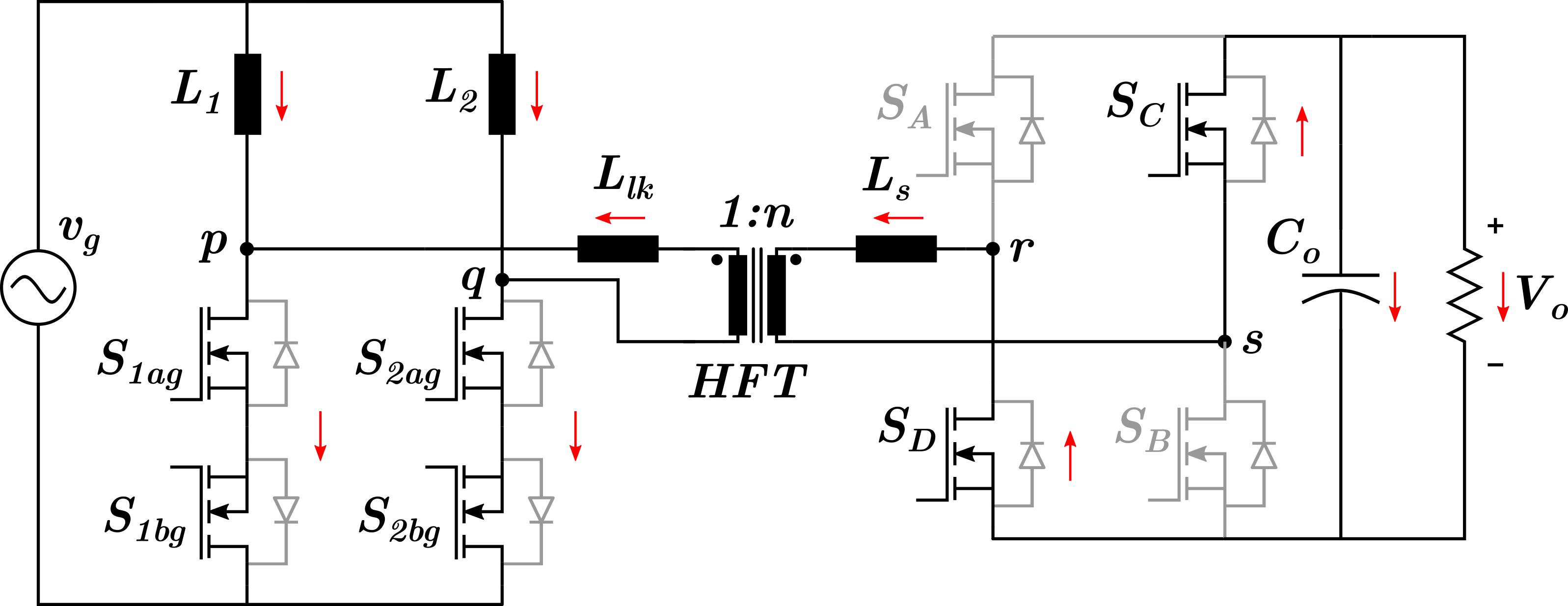}}
          \hfill
     \subfloat [\label{fig:t2-t3}]{
         \centering
         \includegraphics[width=0.32\linewidth]{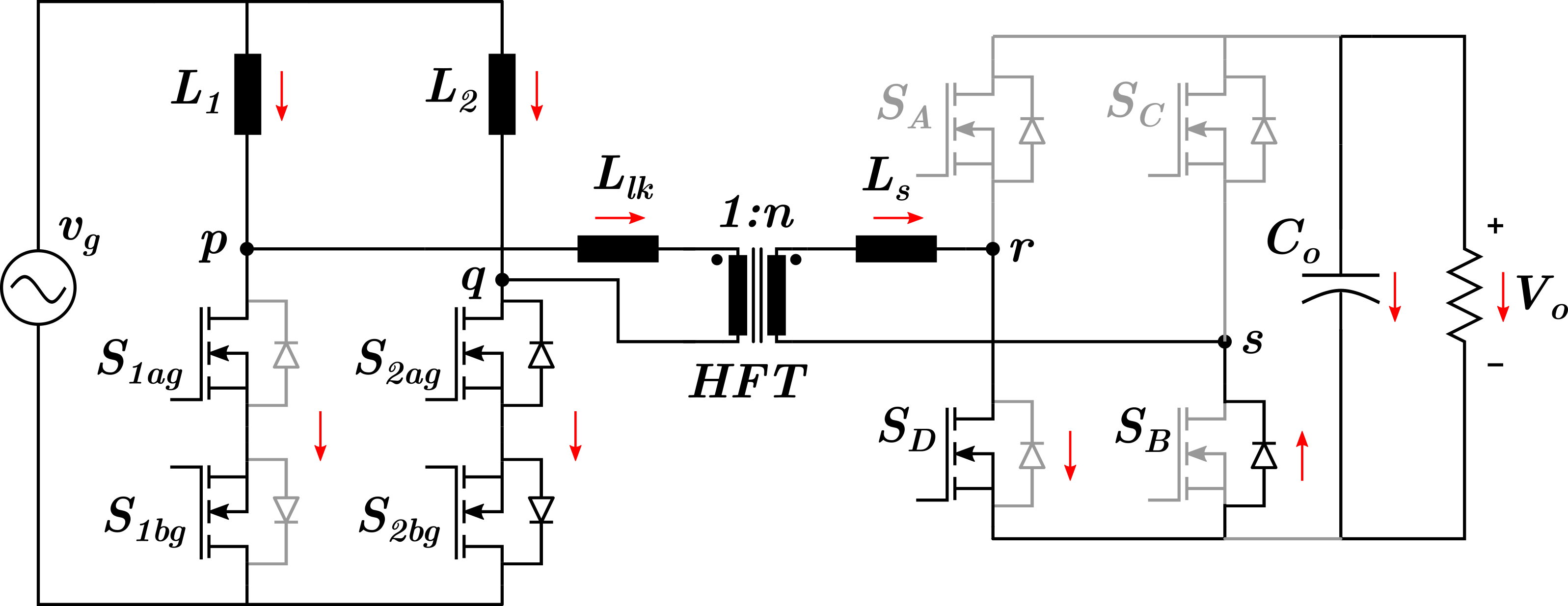}}
          \hfill
     \subfloat [\label{fig:t3-t4}]{
         \centering
         \includegraphics[width=0.32\linewidth]{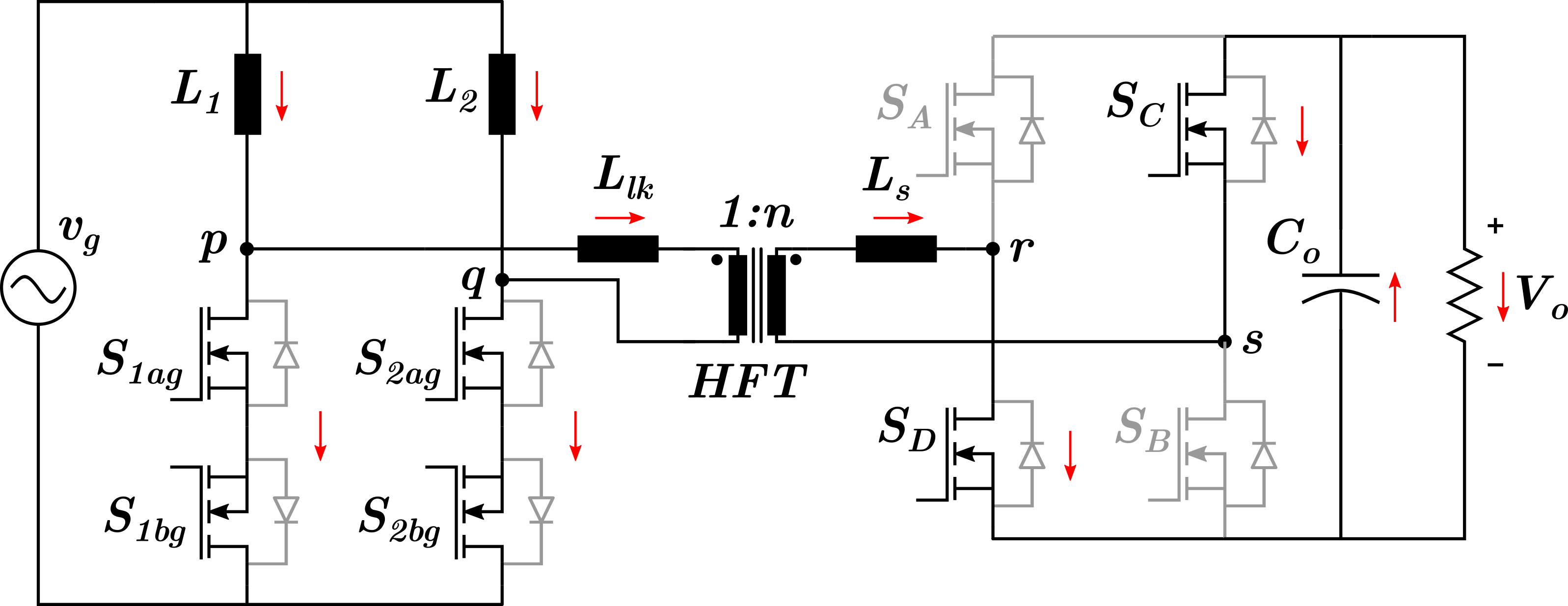}}
         \hfill
     \subfloat [\label{fig:t4-t5}]{
         \centering
         \includegraphics[width=0.32\linewidth]{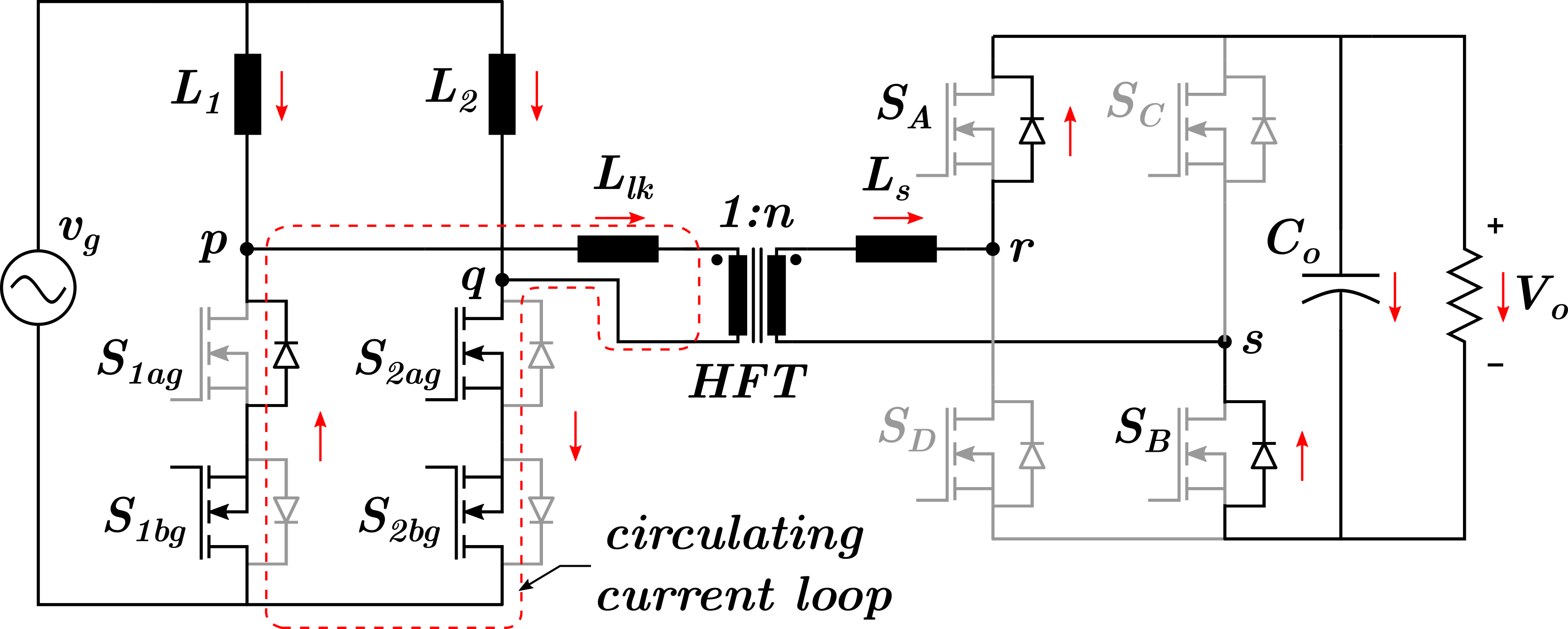}}
          \hfill
     \subfloat [\label{fig:t5-t6}]{
         \centering
         \includegraphics[width=0.32\linewidth]{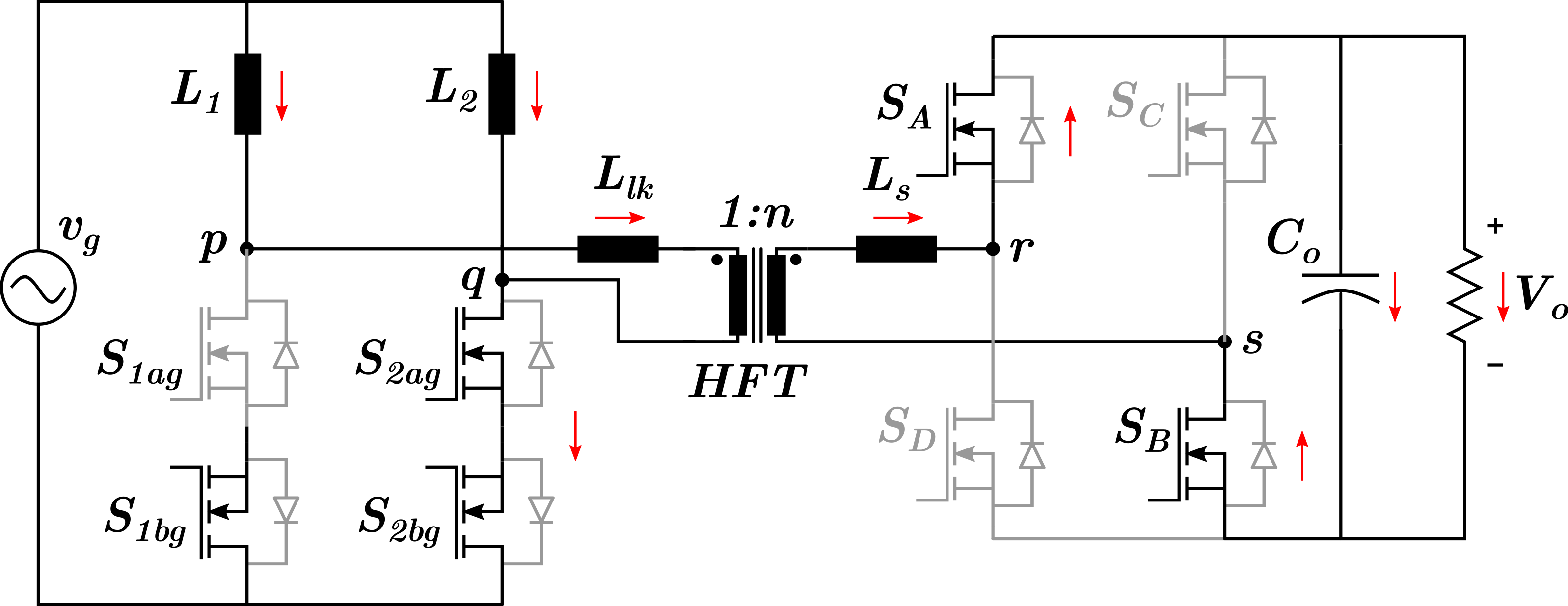}}
         
        \caption{Equivalent circuits of ICFHB ac-dc converter in a switching interval at various stages (a) Stage 1 ($t_0,t_1$), (b) Stage 2 ($t_1,t_2$), (c) Stage 3 ($t_2,t_3$), (d) Stage 4 ($t_3,t_4$), (e) Stage 5 ($t_4,t_5$), (f) Stage 6 ($t_5,t_6$).}
\end{figure*}

\textit{Stage 1 ($t_0,t_1$):} Fig. \ref{fig:t0-t1} shows the equivalent circuit during this stage. At $t_0$, the HFT ac side winding current $i_{lk}^{(k)}(t)$ equals boost inductor current $i_{L2}^{(k)}(t)$. The current through $S_{2ag}$ body diode reaches zero and naturally turns-off achieving ZCS. The output parasitic capacitance of $S_{2ag}$ charges to $V_o/n$, therefore $v_{pq}^{(k)} = -V_o/n$.  $S_{1ag}$ is on in this stage. $S_{C}$ and $S_{D}$ are turned on at $t_0$ and their respective body diode current transfers to the channel of the devices thereby achieving ZVS. $S_{A}$ and $S_{B}$ are off during this stage and their respective device output capacitance are charged to $V_o$. The equation of $i_{lk}^{(k)}(t)$ during stage 1 is given by (\ref{eq:ilkt0_1}) and at $t_0$ is given by (\ref{eq:ilkt0}). The equation of ac side switch current and dc side switch current during this stage is give by (\ref{eq:is1g_0}) and (\ref{eq:scd_0}) respectively.     
\begin{align}
i_{lk}^{(k)}(t)&=-\frac{(v_g^{(k)}-\frac{V_o}{n})}{L_2+L_{t}}(t-t_0)+i_{lk}^{(k)}(t_0)\label{eq:ilkt0_1}\\
i_{lk}^{(k)}(t_0) &= -\frac{1}{2}(i_g^{(k)}+\frac{v_g^{(k)}}{L_2}d_1^{(k)}T_s) \label{eq:ilkt0}\\
i_{S_{1ag}}^{(k)}(t) &= i_{L1}^{(k)}(t)+i_{lk}^{(k)}(t)\label{eq:is1g_0}\\
i_{S_{C}}^{(k)}(t) &= i_{S_{D}}^{(k)}(t)  = \frac{i_{lk}^{(k)}(t)}{n}\label{eq:scd_0} 
\end{align}

\textit{Stage 2 ($t_1,t_2$):} At $t_1$, $S_{2ag}$ is turned on, as a result, its device output capacitance completely discharges in short duration. $S_{1ag}$ is already on from stage 1, therefore $v_{pq}^{(k)} = 0$. $S_{C}$ and $S_{D}$ is also kept on from stage 1 and thus $v_{rs}^{(k)}=-V_o$. Fig. \ref{fig:t1-t2} shows the circuit condition during this stage.  The ac side HFT winding current gets transferred to switch $S_{2ag}$ at a constant slope defined by the total series inductor $L_t$ and $v_{rs}^{(k)}$. The equation of $i_{lk}^{(k)}(t)$ during this stage is given by (\ref{eq:ilkt1_2}) and (\ref{eq:ilkt1}) gives the value of $i_{lk}^{(k)}(t)$  at $t_1$. AC side leg 1 switch current follows the same equation as (\ref{eq:is1g_0}) and leg 2 switch current is as per (\ref{eq:is2g_1}). DC side switch current in stage 2 is same as in stage 1.    
\begin{align}
i_{lk}^{(k)}(t)&=\frac{V_o}{nL_{t}}(t-t_1)+i_{lk}^{(k)}(t_1)\label{eq:ilkt1_2}\\
i_{lk}^{(k)}(t_1) &= -\frac{1}{2}(i_g^{(k)}-\frac{v_g^{(k)}}{L_2}d_1^{(k)}T_s) \label{eq:ilkt1}\\
i_{S_{2ag}}^{(k)}(t) &= -i_{lk}^{(k)}(t)\label{eq:is2g_1}
\end{align}

The duration of this stage is denoted as $\alpha^{(k)}T_s$ as shown in Fig. \ref{fig:90} and is given by
\begin{align}
\alpha^{(k)}T_s=\frac{nL_{t}|i_{lk}^{(k)}(t_1)|}{V_o}\label{eq:alpha}
\end{align}

 \begingroup
\setlength{\tabcolsep}{6pt} 
\renewcommand{\arraystretch}{3}
\begin{table*}[!htbp]
\centering
\caption{Current Expressions}
\label{current}
\begin{tabular}{c|c|c|c}
\hline
\hline
Parameters & SPSM \cite{rathoreAnalysisDesignExperimental2013} & DCPSM\cite{prasannaNovelZeroCurrentSwitchingCurrentFed2013} & Proposed IDCPSM \\
\hline
$i_{lk,rms}^{(k)}$       & \begin{tabular}[c]{@{}c@{}}$\begin{aligned}\frac{i_g^{(k)}}{2}\sqrt{\frac{5}{3}-\frac{4d_{1}^{(k)}}{3}}+i_{lk,pk}^{(k)}\end{aligned}$\\$\begin{aligned}\sqrt{\frac{2d_{1}^{(k)}}{3}-\frac{1}{3}+\frac{i_g^{(k)}}{i_{lk,pk}^{(k)}}\left(\frac{1}{6} -\frac{d_{1}^{(k)}}{3}+\frac{2\beta^{(k)}}{3}\right) }\end{aligned}$\end{tabular}    & \begin{tabular}[c]{@{}c@{}}$\begin{aligned}\frac{i_g^{(k)}}{\sqrt{2}}\sqrt{1-d_{1}^{(k)}+\frac{\alpha^{(k)}}{3}+\frac{\beta^{(k)}}{3}}\end{aligned}$\\$\begin{aligned}+I_{max}\sqrt{\frac{2d_{2}}{3}+\frac{2\beta^{(k)}}{3}+\frac{i_g^{(k)}}{I_{max}}\frac{\beta^{(k)}}{3}}\end{aligned}$\end{tabular}      &  \begin{tabular}[c]{@{}c@{}}$\begin{aligned}\frac{i_g^{(k)}}{\sqrt{2}}\end{aligned}\sqrt{\begin{aligned}1&-d_{1}^{(k)}+\frac{4d_{2}^{(k)}}{3}\\&+\frac{\alpha^{(k)}}{3}+\frac{7\beta^{(k)}}{3}\end{aligned}}$ \end{tabular}        \\
\hline
$i_{S_{xyg,rms}}^{(k)}$      &  \begin{tabular}[c]{@{}c@{}}$\begin{aligned}\frac{i_g^{(k)}}{2}\sqrt{\frac{5}{3}-\frac{4d_{1}^{(k)}}{3}}+i_{lk,pk}^{(k)}\end{aligned}$\\$\begin{aligned}\sqrt{\frac{2d_{1}^{(k)}}{3}-\frac{1}{3}-\frac{\beta^{(k)}}{3}+\frac{i_g^{(k)}}{i_{lk,pk}^{(k)}}\left(\frac{\beta^{(k)}}{3}-\frac{d_{1}^{(k)}}{3}-\frac{1}{6}\right) }\end{aligned}$\end{tabular}        &   \begin{tabular}[c]{@{}c@{}}$\begin{aligned}\frac{i_g^{(k)}}{2}\sqrt{\frac{11}{3}-\frac{10d_{1}^{(k)}}{3}+\frac{4d_{2}^{(k)}}{3}+\frac{7\alpha^{(k)}}{3}+\frac{5\beta^{(k)}}{3}}\end{aligned}$\\$\begin{aligned}+I_{max}\sqrt{\frac{2d_{2}}{3}+\frac{\beta^{(k)}}{3}+\frac{i_g^{(k)}}{I_{max}}\frac{2\beta^{(k)}}{3}}\end{aligned}$\end{tabular}    &   \begin{tabular}[c]{@{}c@{}}$\begin{aligned}\frac{i_g^{(k)}}{2}\end{aligned}\sqrt{\begin{aligned}\frac{11}{3}&-\frac{10d_{1}^{(k)}}{3}+4d_{2}^{(k)}\\&+\frac{7\alpha^{(k)}}{3}+\frac{17\beta^{(k)}}{3}\end{aligned}}$ \end{tabular}        \\
\hline
$i_{D_{xyg,avg}}^{(k)}$      &  \begin{tabular}[c]{@{}c@{}}$\begin{aligned}\left(i_{lk,pk}^{(k)}-\frac{i_g^{(k)}}{2}\right) \frac{\beta^{(k)}}{2} \end{aligned}$\end{tabular}   &\begin{tabular}[c]{@{}c@{}}$\begin{aligned}\left(I_{max}-\frac{i_g^{(k)}}{2}\right) \frac{\beta^{(k)}}{2} \end{aligned}$\end{tabular}      & \begin{tabular}[c]{@{}c@{}}$\begin{aligned}\frac{i_g^{(k)}}{2}\frac{\beta^{(k)}}{2} \end{aligned}$\end{tabular}           \\
\hline
$i_{S_{p,rms}}^{(k)}$       &\begin{tabular}[c]{@{}c@{}}$\begin{aligned}\frac{i_g^{(k)}}{2n}\sqrt{\frac{5}{6}-\frac{2d_{1}^{(k)}}{3}}+i_{lk,pk}^{(k)}\end{aligned}$\\$\begin{aligned}\sqrt{\frac{d_{1}^{(k)}}{3}-\frac{1}{6}+\frac{i_g^{(k)}}{i_{lk,pk}^{(k)}}\left(\frac{1}{12} -\frac{d_{1}^{(k)}}{6}+\frac{\beta^{(k)}}{3}\right) }\end{aligned}$\end{tabular}      &   \begin{tabular}[c]{@{}c@{}}$\begin{aligned}\frac{I_{max}}{n}\sqrt{\frac{d_{2}}{3}}\end{aligned}$\end{tabular}   & \begin{tabular}[c]{@{}c@{}}$\begin{aligned}\frac{i_g^{(k)}}{2n}\end{aligned}\sqrt{\begin{aligned}1&-d_{1}^{(k)}+\frac{4d_{2}^{(k)}}{3}\\&+\frac{\alpha^{(k)}}{3}+\frac{7\beta^{(k)}}{3}\end{aligned}}$ \end{tabular}           \\
\hline
$i_{D_{p,avg}}^{(k)}$       &   0  & \begin{tabular}[c]{@{}c@{}}$\begin{aligned}\frac{i_g^{(k)}}{2n}\left( 1-d_{1}^{(k)}+\frac{\alpha^{(k)}}{2}+\frac{\beta^{(k)}}{2}\right)+\frac{I_{max}}{n}\frac{\beta^{(k)}}{2} \end{aligned}$\end{tabular}      &    0      \\
\hline
\hline
\end{tabular}
\end{table*}
\endgroup

\begin{figure*}[!htbp]
     \centering
     \subfloat[\label{fig:i_lk_g}]
	{
         \centering
         \includegraphics[width=0.235\linewidth]{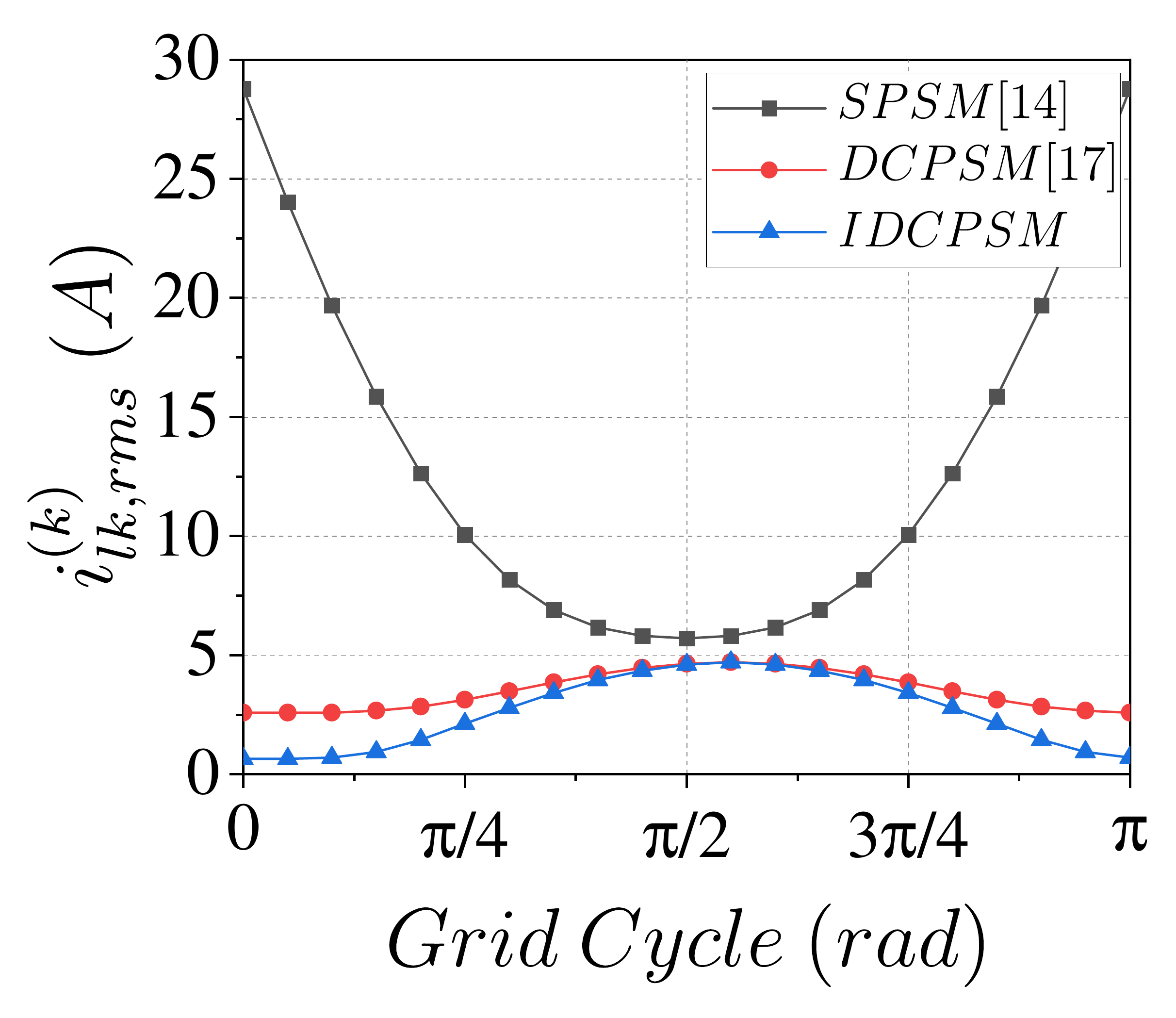}
	}     \hfill
     \subfloat[\label{fig:i_s1_g}]
	{
         \centering
         \includegraphics[width=0.235\linewidth]{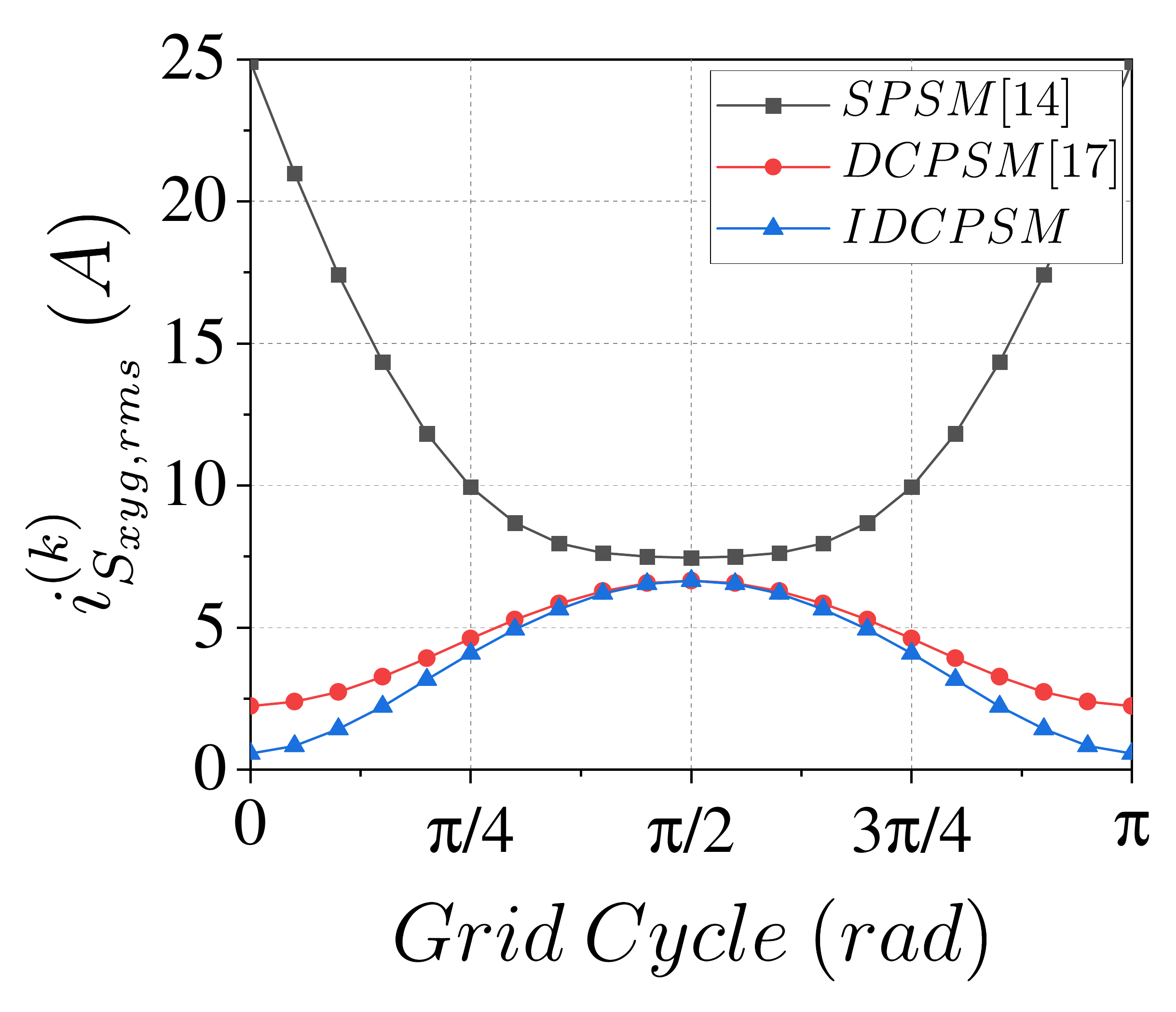}
	}     \hfill
     \subfloat[\label{i_sa_g}]
	{
         \centering
         \includegraphics[width=0.235\linewidth]{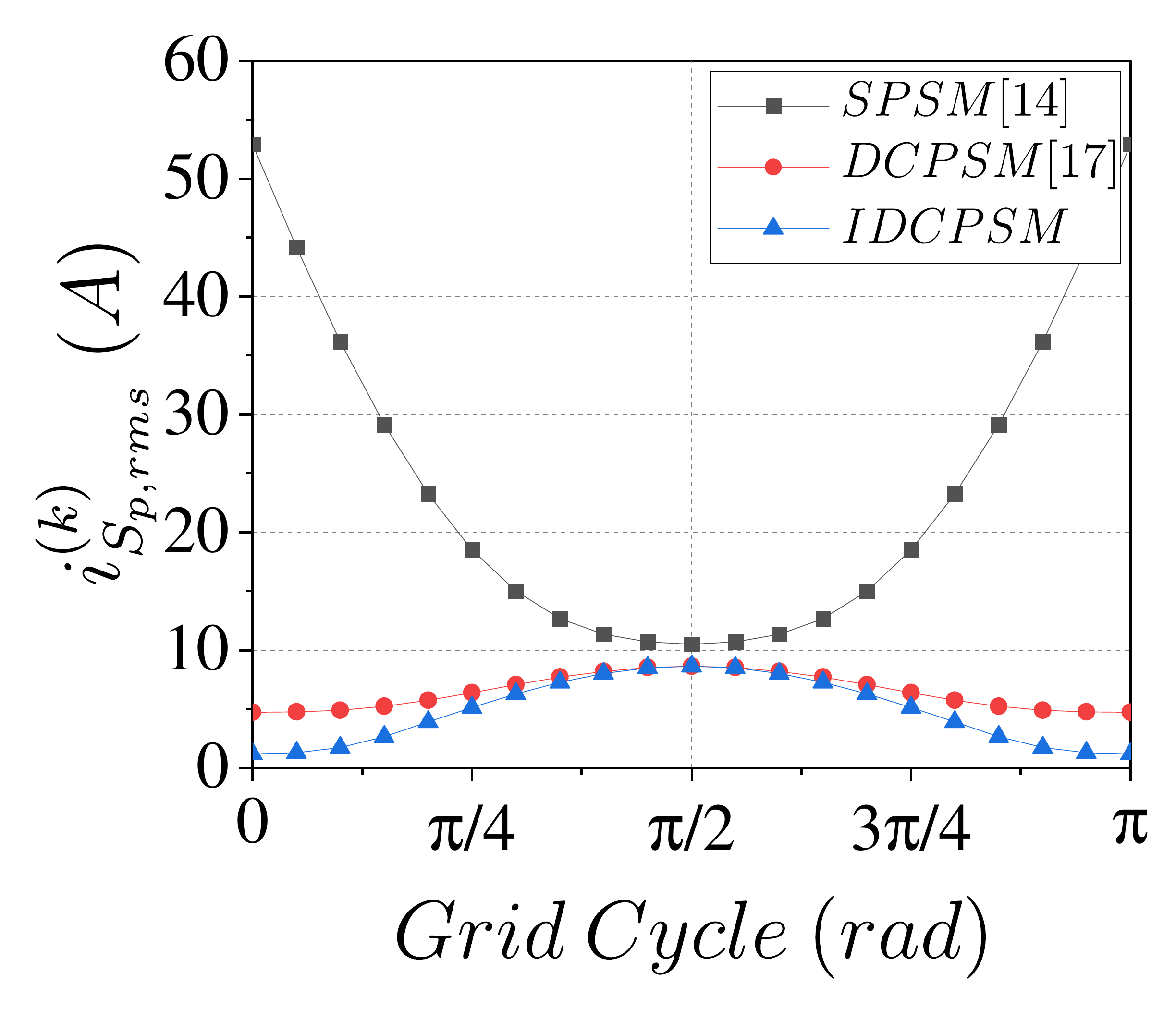}
	}
	  \subfloat[\label{i_d1_g}]
	{
         \centering
         \includegraphics[width=0.235\linewidth]{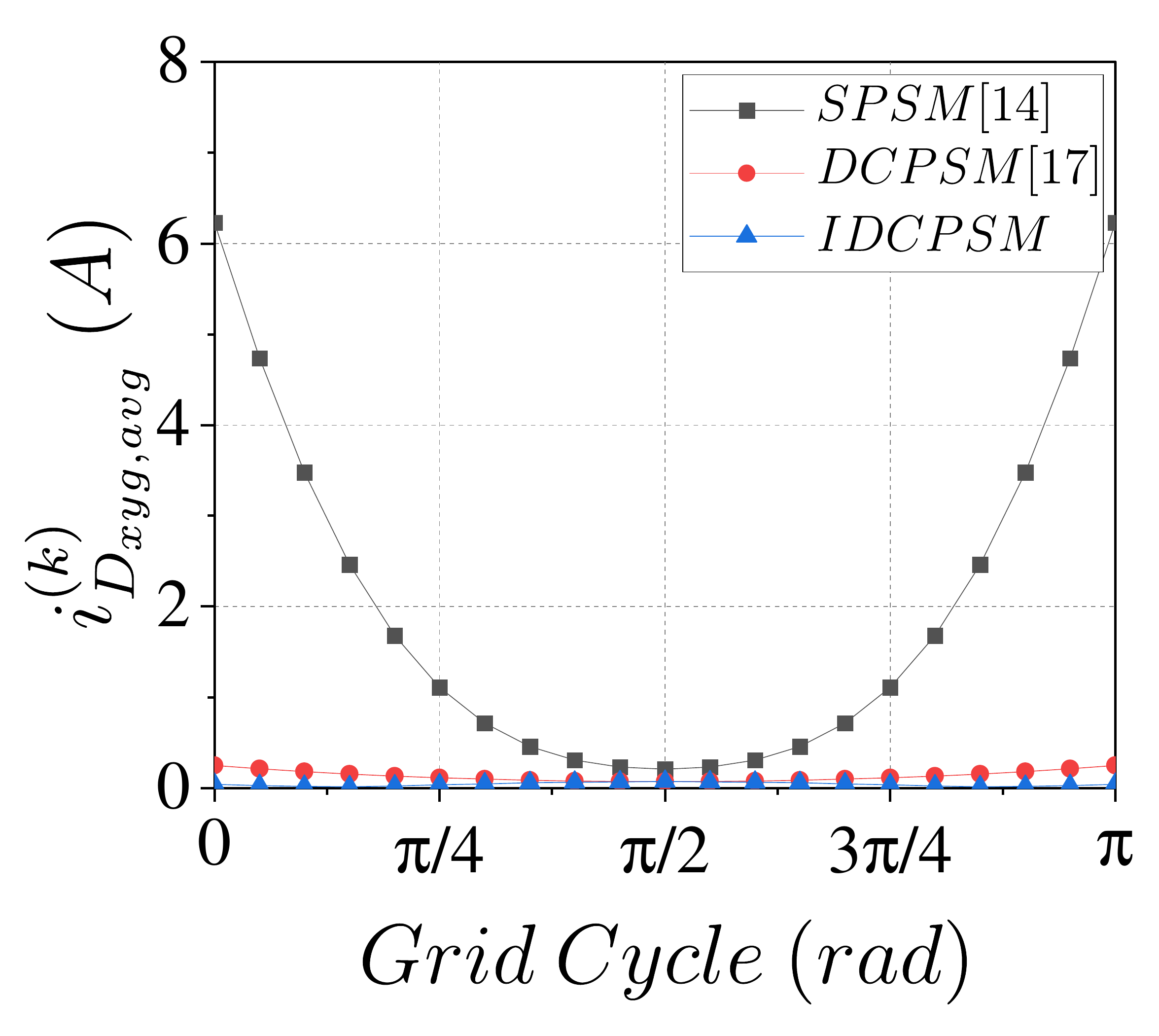}

	}

        \caption{Variation of switching cycle RMS and average value of current in a half grid cycle for different modulation scheme at $P_o = 1.5\:kW$ and $V_o = 345\:V$ (a) RMS current through the ac side winding of HFT, (b) RMS current of ac side switch, (c) RMS current of dc side switch, (d) average current of ac side body diode.}
        \label{current_plot}
\end{figure*}

\textit{Stage 3 ($t_2,t_3$):} At $t_2$, HFT winding current $i_{lk}$ reaches zero as the line inductor current $i_{L2}$ gets completely  transferred to switch $S_{2ag}$. Since both $S_{1ag}$ and $S_{2ag}$ is on, $v_{pq}^{(k)} = 0$. $S_{D}$ is kept on and gate pulse to $S_{C}$ is removed. Since $v_{rs}^{(k)} =-V_o$ at $t_2$, this voltage is kept applied across the winding of HFT for a short duration even after $t_2$. This rises the HFT winding current to a minimum value which in turn discharges the output device capacitance of $S_{B}$ and charges the the output device capacitance of $S_{C}$. The body diode of $S_{B}$ gets forward biased and the HFT winding current is maintained at that minimum value as $v_{rs}^{(k)} = 0$. This minimum current value is very small compared to the total load current and hence it is assumed to be zero. Fig. \ref{fig:t2-t3} shows the circuit condition during this stage. 

\textit{Stage 4 ($t_3,t_4$):} Fig. \ref{fig:t3-t4} shows the equivalent circuit during this stage.  $S_{1ag}$ and $S_{2ag}$ is kept on and $v_{pq}^{(k)} = 0$. At $t_3$, $S_{C}$ is turned on and its output device capacitance discharges and at the same time output device capacitance of $S_{B}$ charges to $V_o$. Thus, $v_{rs}^{(k)} = -V_o$ and is applied across the dc side winding of HFT, rising the current through it at a constant slope. The current through $S_{1ag}$ falls at the same constant slope and through $S_{2ag}$ rises at the same constant slope. This duration is kept sufficiently long enough for the current to go negative through $S_{1ag}$ as shown in Fig. \ref{fig:10_current}. The negative current flowing through the loop of ac side switches and the HFT is the circulating current $i_{cir}^{(k)}$ as shown in Fig. \ref{fig:t4-t5}. The equation of $i_{lk}^{(k)}(t)$ during this stage is given by (\ref{eq:ilkt3_4}). AC side leg 1 switch current and leg 2 switch current is as per (\ref{eq:is1g_4}) and (\ref{eq:is2g_4}) respectively. DC side switch current equation in stage 4 is same as in stage 1.    
\begin{align}
i_{lk}^{(k)}(t)&=\frac{V_o}{nL_{t}}(t-t_3)\label{eq:ilkt3_4}\\
i_{S_{1ag}}^{(k)}(t) &= i_{L1}^{(k)}(t)-i_{lk}^{(k)}(t)\label{eq:is1g_4}\\
i_{S_{2ag}}^{(k)}(t) &= i_{L2}^{(k)}(t)+i_{lk}^{(k)}(t)\label{eq:is2g_4}
\end{align}
From Fig. \ref{fig:90} it can be noted that the duration of this stage is the secondary side duty $d_2^{(k)}T_s$ and is given by
\begin{align}
d_2^{(k)}T_s=\frac{nL_{t}|i_{lk,pk}^{(k)}|}{V_o}\label{eq:duty2}
\end{align}
It is clear from (\ref{eq:duty2}) that the peak value of ac side HFT winding current $i_{lk,pk}^{(k)}$ is dependent on $d_2^{(k)}$. In the existing modulation scheme, the value of $d_2^{(k)}$ was decided for the maximum output power and was kept constant irrespective of the operating conditions \cite{prasannaNovelZeroCurrentSwitchingCurrentFed2013}. Thus, $i_{lk,pk}=I_{max}$ remained constant across the entire grid cycle of ac mains voltage and in different load conditions. This can be observed from Fig. \ref{fig:90} that at the peak of ac mains voltage i.e at $\omega \tau=90^{\circ}$, the winding current reaches $I_{max}$ during $d_2T_s$. For this fixed value of $d_2^{(k)}$, the peak current value remains constant near the zero crossing of ac mains voltage i.e at $\omega \tau=10^{\circ}$ as shown by dotted lines $(t_4'-t_3')T_s$ in Fig. \ref{fig:10}. Fig. \ref{fig:10_current} shows the current through the ac side and dc side switches near the zero crossing. It can be observed in the figure, as shown by dotted lines, that the peak current through the switches remains constant across the grid cycle. From (\ref{eq:cir}), it is evident that for a fixed value of $i_{lk,pk}=I_{max}$, the circulating current flowing through the converter is substantially high near the zero crossing of ac mains voltage. This high circulating current attributes to increased conduction losses in both ac and dc side switches and in the HFT.
\begin{align}
i_{cir}^{(k)}=i_{lk,pk}^{(k)}-\frac{1}{2}i_{g}^{(k)}\label{eq:cir}
\end{align}
  
In the proposed IDCPSM, the ratio between $i_g^{(k)}$ and $i_{lk,pk}^{(k)}$ is kept fixed in the entire grid cycle of ac mains voltage i.e $i_{lk,pk}^{(k)}$ follows the grid current sinusoidal envelope. This is achieved by varying $d_2^{(k)}$ in every switching interval. This can observed from Fig. \ref{fig:10} that near the zero crossing of ac mains voltage, $d_2^{(k)}$, is adjusted as per the fixed ratio between $i_g^{(k)}$ and $i_{lk,pk}^{(k)}$. This substantially reduces the peak current value through the HFT winding as shown by bold lines. The merits of the proposed IDCPSM can also be seen in Fig. \ref{fig:10_current}. The peak current through the ac and dc side switches has also reduced substantially as shown by bold lines near the zero crossing of ac mains voltage. As per (\ref{eq:cir}) the circulating current will also follow the sinusoidal envelope across the grid cycle. This reduced peak and circulating currents through the switches and the HFT contributes to lower conduction losses, significantly improving the efficiency of the converter.   

\textit{Stage 5 ($t_4,t_5$):} At $t_4$, $S_{1ag}$ is turned off and the negative current flowing through it shifts to its body diode. $S_{2ag}$ continues to stay on from the previous stage therefore $v_{pq}^{(k)} = 0$. $S_{C}$ and $S_{D}$ are turned off and its output parasitic capacitance charges to $V_o$ and at the same time output parasitic capacitance of $S_{A}$ and $S_{B}$ discharges and their body diode starts conducting. Thus, $v_{rs}^{(k)} = V_o$ and is applied across the dc side winding of HFT and current through it falls at a constant slope. The equation of $i_{lk}^{(k)}(t)$ during this stage is given by (\ref{eq:ilkt4_5}) at $t_4$ is given by (\ref{eq:ilkt4}). AC side switch current follows the same equation as per stage 4. DC side body diode current equation is given by (\ref{eq:dab_4}).    
\begin{align}
i_{lk}^{(k)}(t)&=\frac{-V_o}{nL_{t}}(t-t_4)+i_{lk}^{(k)}(t_4)\label{eq:ilkt4_5}\\
i_{lk}^{(k)}(t_4) &= \frac{V_od_2^{(k)}T_s}{nL_t} = i_{lk,pk}^{(k)}\label{eq:ilkt4}\\
i_{D_{A}}^{(k)}(t) &= i_{D_{B}}^{(k)}(t)  = -\frac{i_{lk}^{(k)}(t)}{n}\label{eq:dab_4} 
\end{align}
As represented in Fig. \ref{fig:90}, this duration is $\beta T_s$ and is given by
\begin{align}
\beta^{(k)}T_s=\frac{nL_{t}|i_{lk}^{(k)}(t_5)-i_{lk}^{(k)}(t_4)|}{V_o}\label{eq:beta}
\end{align}

\textit{Stage 6 ($t_5,t_6$):} At $t_5$, the current flowing through the body diode of $S_{1ag}$ goes to zero, naturally turning off the device and achieving ZCS. $S_{2ag}$ continues to stay on therefore $v_{pq}^{(k)} = V_o/n$. $S_{A}$ and $S_{B}$ are turned on and the current through their body diode shifts to channel undergoing ZVS. $v_{rs}^{(k)} = V_o$. The equation of $i_{lk}^{(k)}(t)$ during this stage is given by (\ref{eq:ilkt5_6}) and at $t_5$ is given by (\ref{eq:ilkt5}). AC side leg 2 switch current equation in given by (\ref{eq:is1g_5}) and dc side switch current equation is given by (\ref{eq:sab_5}).    
\begin{align}
i_{lk}^{(k)}(t)&=\frac{(v_g^{(k)}-\frac{V_o}{n})}{L_1+L_{t}}(t-t_5)+i_{lk}^{(k)}(t_5)\label{eq:ilkt5_6}\\
i_{lk}^{(k)}(t_5) &= \frac{1}{2}(i_g^{(k)}+\frac{v_g^{(k)}}{L_1}d_1^{(k)}T_s) \label{eq:ilkt5}\\
i_{S_{2ag}}^{(k)}(t) &= i_{L2}^{(k)}(t)+i_{lk}^{(k)}(t)\label{eq:is1g_5}\\
i_{S_{A}}^{(k)}(t) &= i_{S_{B}}^{(k)}(t)  = -\frac{i_{lk}^{(k)}(t)}{n}\label{eq:sab_5} 
\end{align}

 Stage 1 to stage 5 describes the converter operation in a half switching interval. The other half is symmetrical to the above described stages as evident from stage 6. 
 
 Table \ref{current} presents the expressions of rms current through ac side HFT winding $i_{lk,rms}^{(k)}$, rms current through ac side switch $i_{S_{xyg,rms}}^{(k)}$ and dc side switch $i_{S_{p,rms}}^{(k)}$, average current through the body diode of ac side switch $i_{D_{xyg,avg}}^{(k)}$  and dc side switch $i_{D_{p,avg}}^{(k)}$ in a switching interval $k$ and Fig. \ref{current_plot} shows their variation in a half grid cycle for SPSM, DCPSM and proposed IDCPSM. Fig. \ref{fig:i_lk_g} shows that the ac side HFT winding rms current is maximum in switching intervals near the zero crossing of ac mains voltage for SPSM and is substantially low for DCPSM. In the proposed IDCPSM, this value is further reduced as the peak current through the windings of HFT is significantly reduced by modulating $d_2^{(k)}$ across the grid cycle. Reduction can also be seen in rms current through ac and dc side switches for the proposed IDCPSM as depicted in Fig. \ref{fig:i_s1_g}-\ref{i_d1_g}. This amounts to lower conduction losses with considerable improvement in the efficiency of the converter.

\begin{figure}[t]
  \centering
    \includegraphics[width=\linewidth]{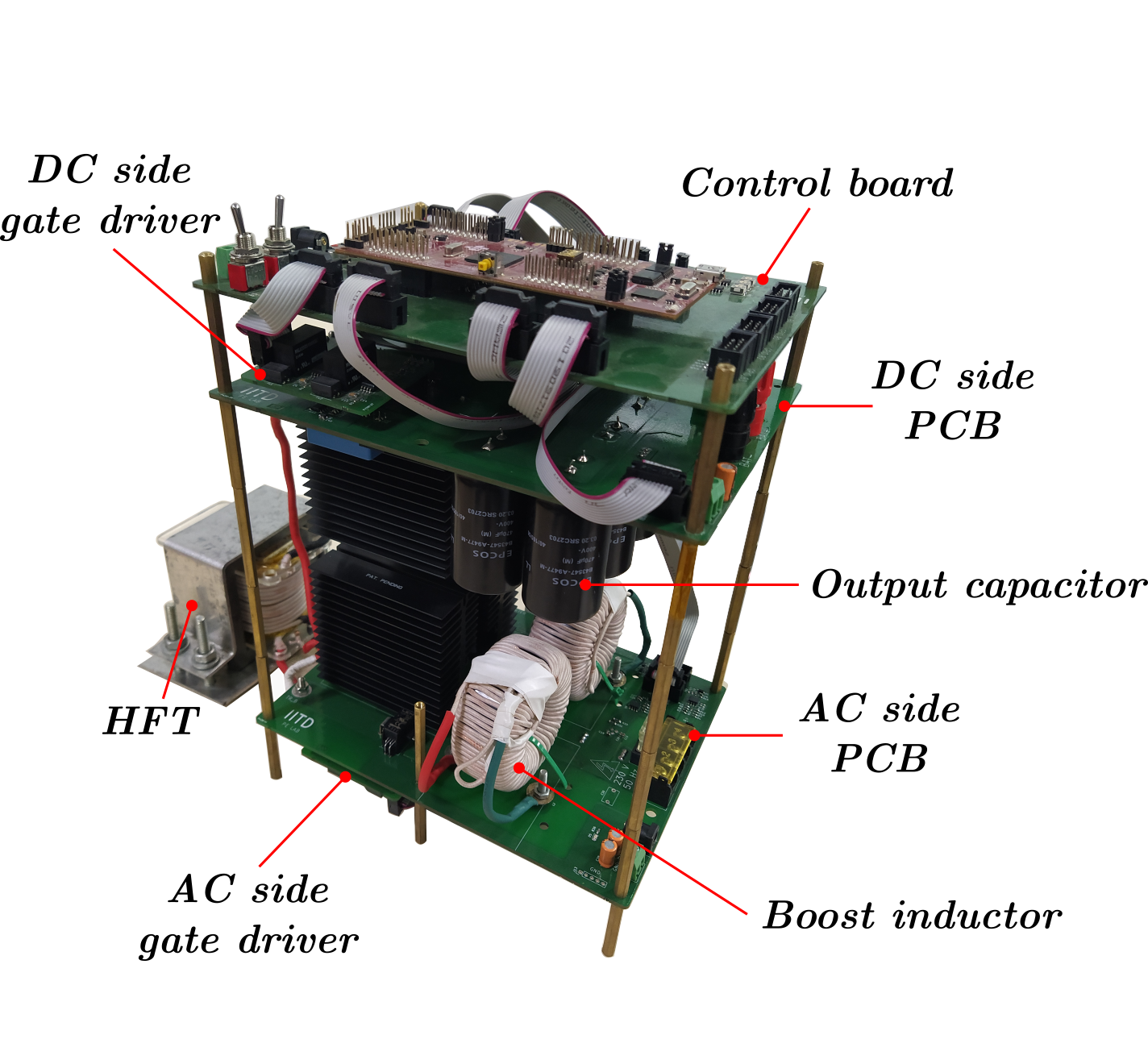}
    \caption{Laboratory prototype of a L-L type ICFHB ac-dc converter.}
        \label{setup}
\end{figure}

\begingroup
\setlength{\tabcolsep}{6pt} 
\renewcommand{\arraystretch}{1.3}
\begin{table}[t]
\caption{Parameters of the implemented hardware prototype}
\label{parameter}
\begin{tabular}{lcc}
\hline
\hline
AC mains voltage    & $V_g$                   & 230 V$_{rms}$                                                                                                                                                                                  \\ \hline
AC mains frequency  & $f_g$                   & 50 Hz                                                                                                                                                                                     \\ \hline
Output Voltage      & $v_o$                   & 300 V...400 V                                                                                                                                                                             \\ \hline
Output Power        & $P_o$                   & 1.5 kW
\\ \hline
Switching Frequency & $f_{sw}$                  & 100 kHz                                                                                                                                                                                  \\ \hline
Transformer         & $HFT$                  & \begin{tabular}[c]{@{}c@{}}Transformer ratio: 0.38 \\ Prim: 26 turns, 135 wires, \#40 SWG\\ Sec: 10 turns, 260 wires, \#40 SWG\\ Core: Ferrite EE80/20 (2 stacked)\\ $L_{lk}$ = 600 nH\end{tabular} \\ \hline
Boost Inductor      & $L_1$, $L_2$               &\begin{tabular}[c]{@{}c@{}}740 $\mu$H \\ 82 turns, 135 wires, \#40 SWG\\ Core: Kool M$\mu$ 77620\end{tabular} \\ \hline
Series Inductor     & $L_s$                   &  \begin{tabular}[c]{@{}c@{}}7.5 $\mu$H  \\ 10 turns, 260 wires, \#40 SWG\\ Core: Kool M$\mu$ 77071\end{tabular} \\ \hline
Output Capacitor    & $C_o$                   & \begin{tabular}[c]{@{}c@{}} 940 $\mu$F\\ 4 parallel leg\\ Each leg: 2 series connected 470 $\mu$F\\ (Electrolytic 400 V)\end{tabular}                                                                       \\ \hline
AC side SiC Mosfets  & $S_{xyg}$ & Cree C2M0080170P                                                                       \\ \hline
DC side SiC Mosfets  & $S_{p}$ & UnitedSiC UF3C065030K4S  \\ \hline
\hline
\end{tabular}
\end{table}
\endgroup

\section{Experimental Results}
A 1.5 kW laboratory prototype was developed and is shown in Fig. \ref{setup}. The performance of the converter with proposed modulation scheme is validated at different loading conditions. Table \ref{parameter} lists the parameters of the implemented prototype. The control scheme is implemented in a TI TMS320F28397D launchpad and the combinational logic of gate pulses is generated using Xilinx XC6SLX4 based FPGA board. 

\begin{figure*}[!htbp]
     \centering
     \subfloat[\label{fig:grid_cycle_old}]
	{
         \centering
         \includegraphics[width=0.31\linewidth]{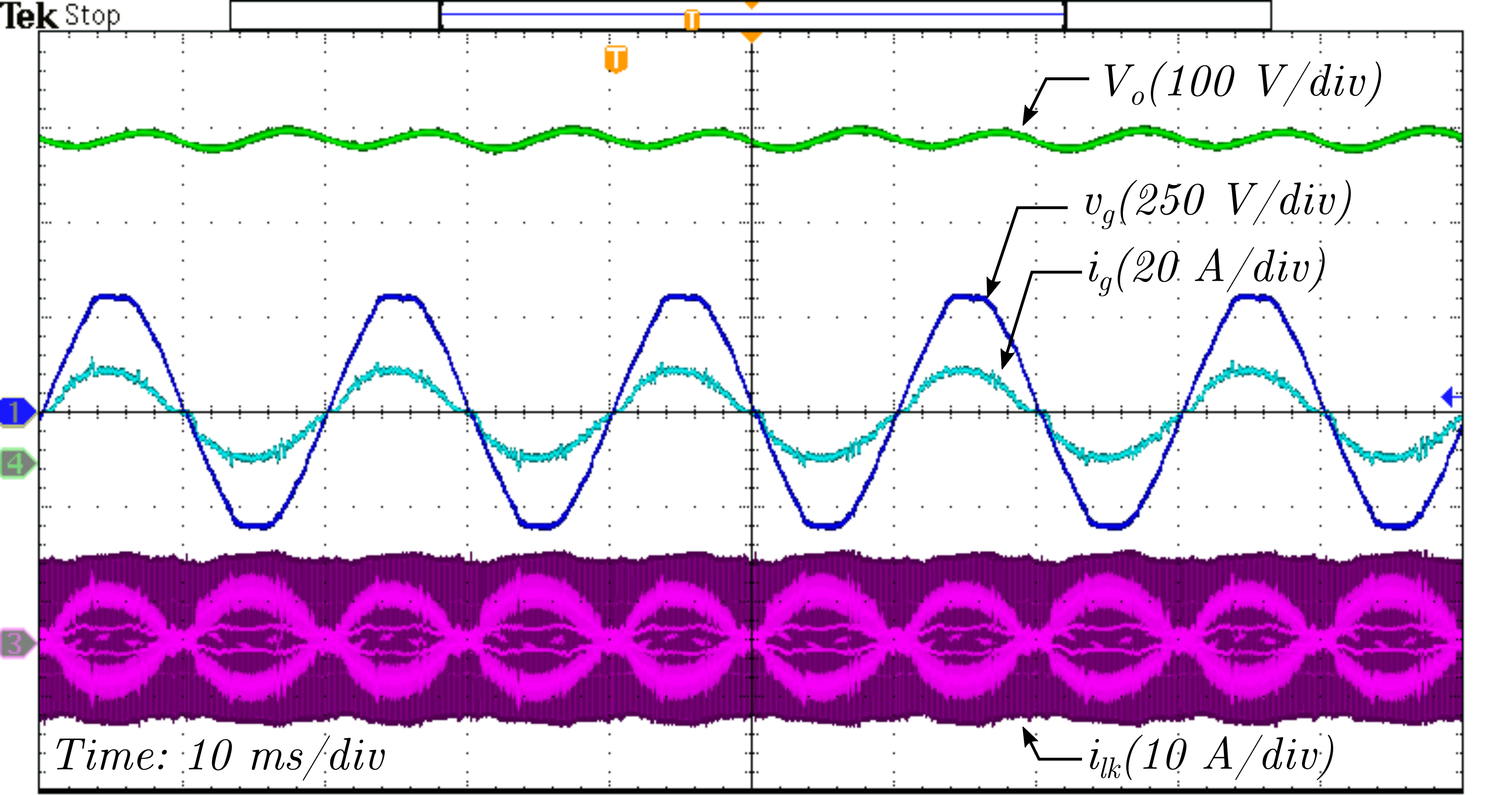}
	}     \hfill
	 \subfloat[\label{fig:peak_zoomed_old}]
	{
         \centering
         \includegraphics[width=0.31\linewidth]{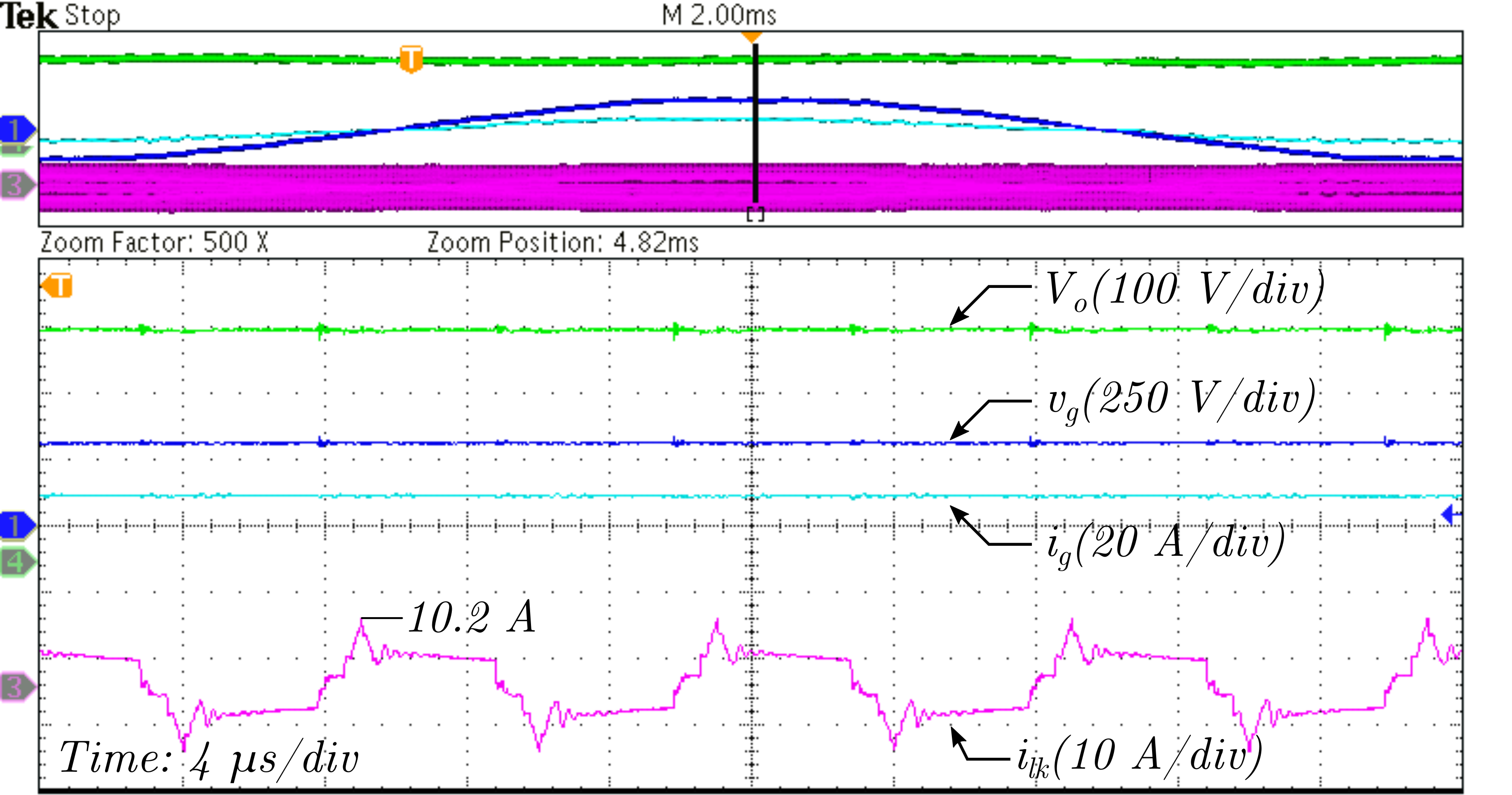}
	}     \hfill
     \subfloat[\label{fig:zero_zoomed_old}]
	{
         \centering
         \includegraphics[width=0.31\linewidth]{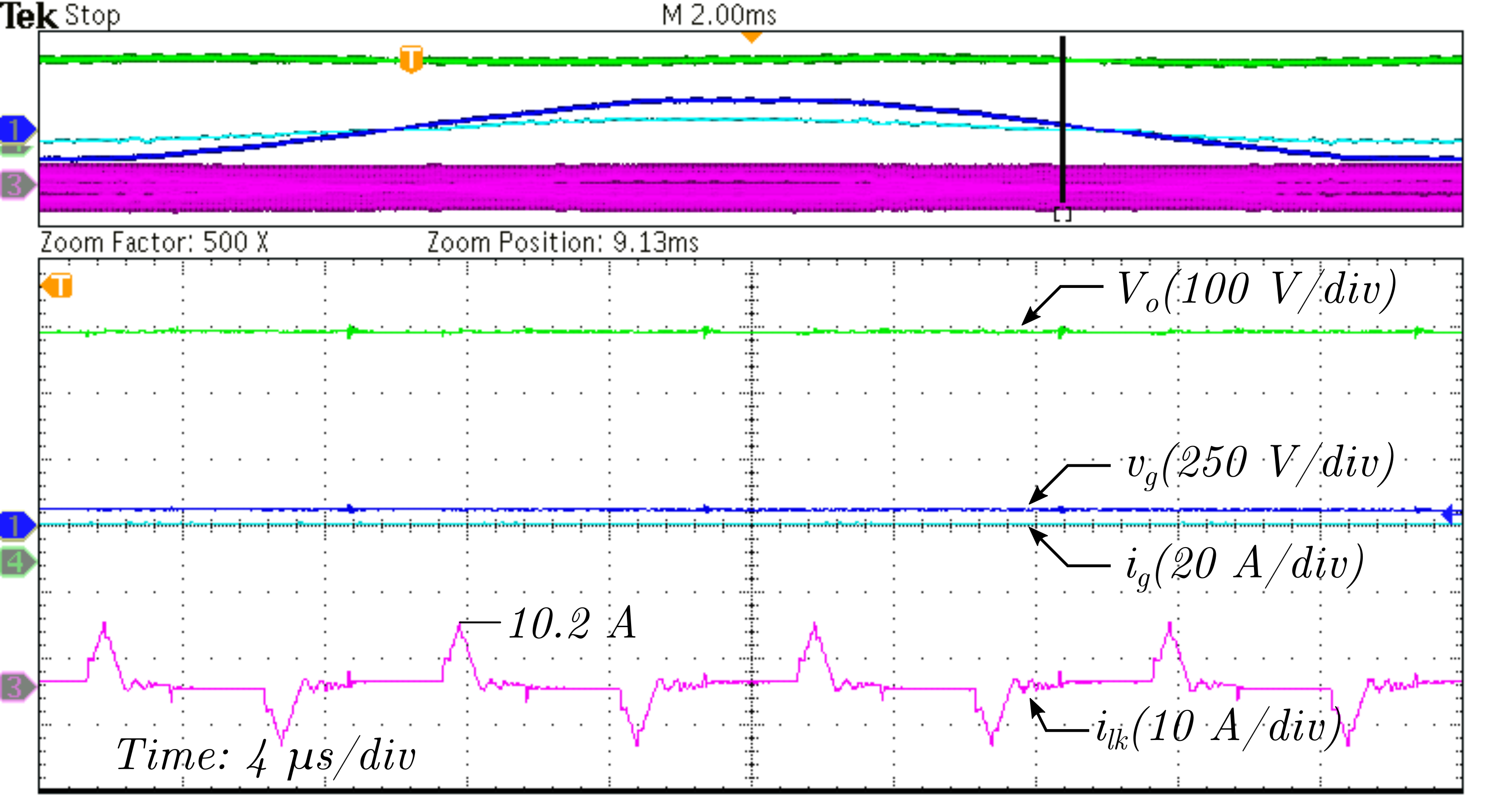}
	}
        \caption{Measured grid voltage $v_g$, grid current $i_g$, output voltage $V_o$ and ac side HFT winding current $i_{lk}$ for DCPSM in (a) grid cycle, (b) zoomed near zero crossing of ac mains voltage. ($P_o = 1.5\:kW, v_g = 230\:V_{rms}$ and $V_o = 345\:V$)}
\end{figure*}

\begin{figure*}[!htbp]
     \centering
     \subfloat[\label{fig:grid_cycle_new}]
	{
         \centering
         \includegraphics[width=0.31\linewidth]{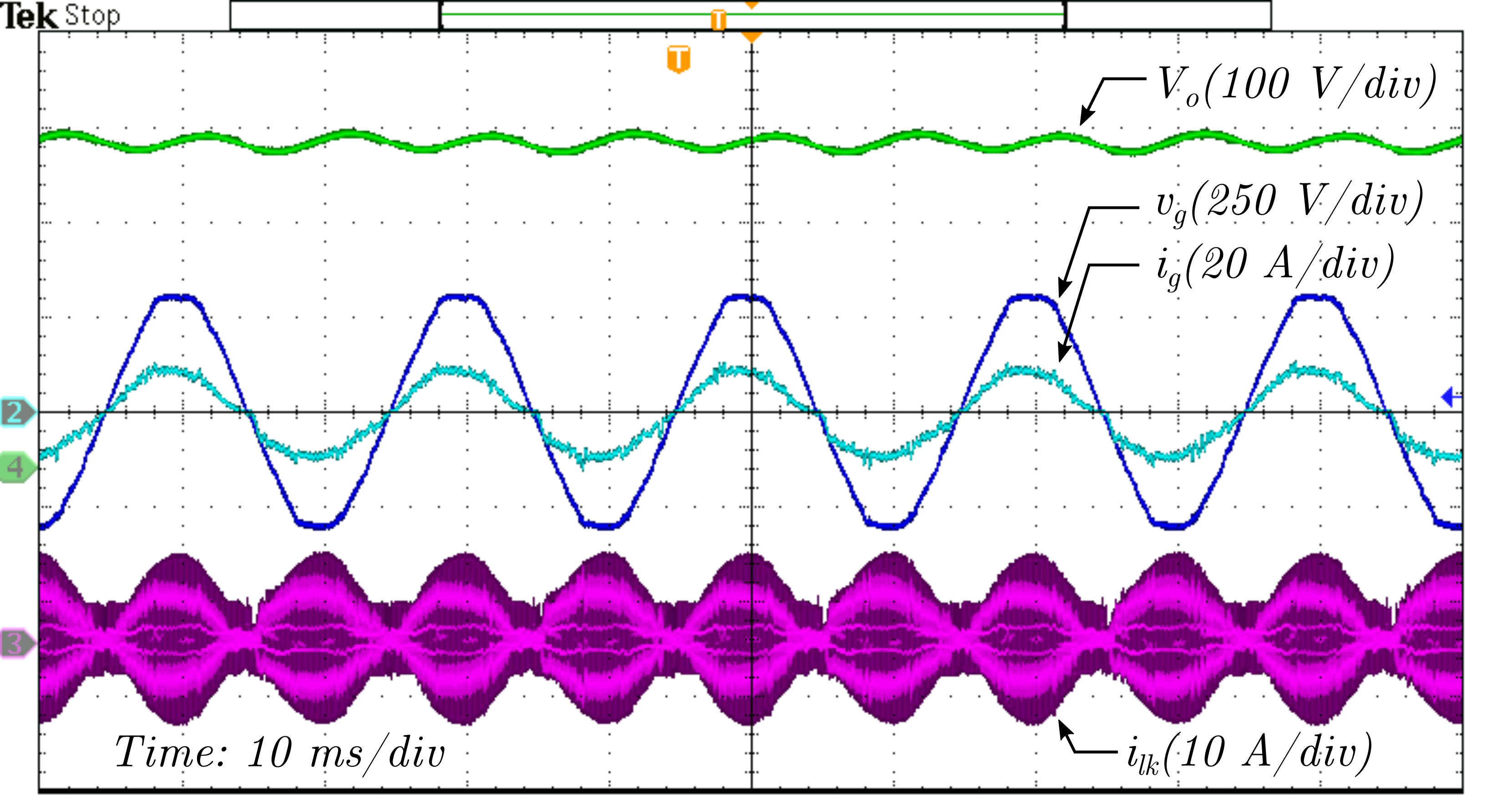}
	}     \hfill
     \subfloat[\label{fig:peak_zoomed_new}]
	{
         \centering
         \includegraphics[width=0.31\linewidth]{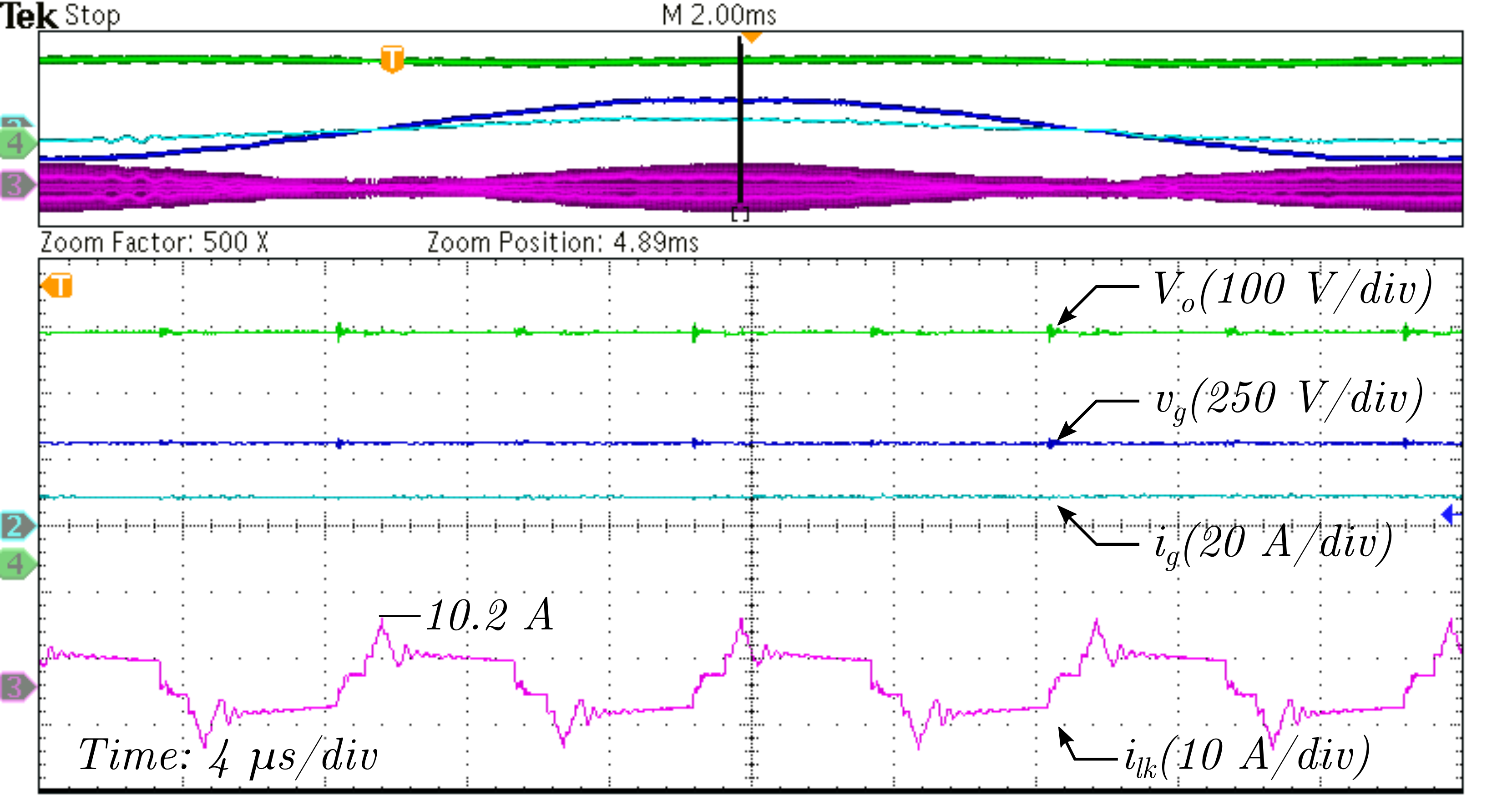}
	}     \hfill
     \subfloat[\label{fig:zero_zoomed_new}]
	{
         \centering
         \includegraphics[width=0.31\linewidth]{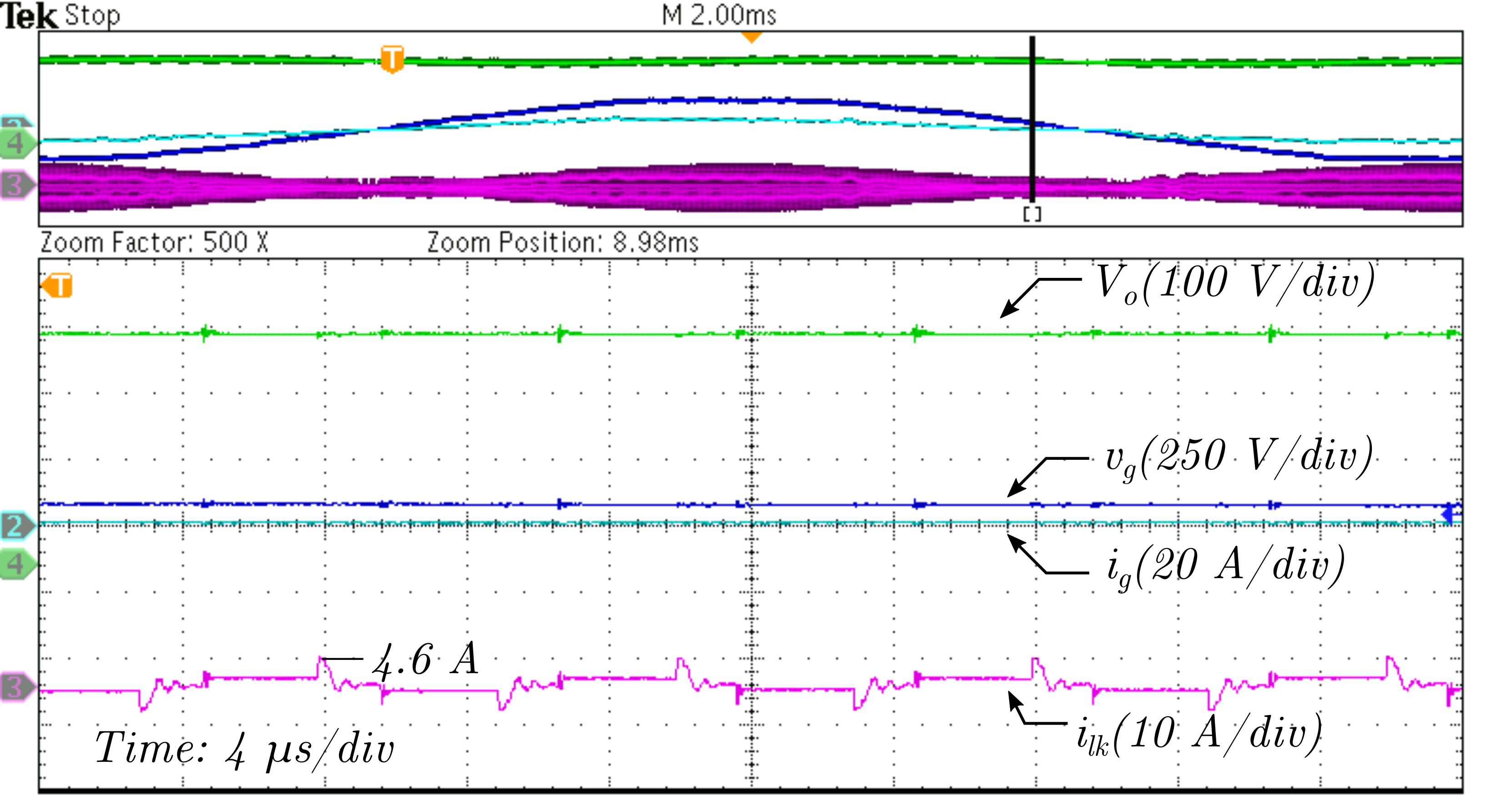}
	}
        \caption{Measured grid voltage $v_g$, grid current $i_g$, output voltage $V_o$ and ac side HFT winding current $i_{lk}$ for IDCPSM in (a) grid cycle, (b) zoomed near peak of ac mains voltage, (c) zoomed near zero crossing of ac mains voltage. ($P_o = 1.5\:kW, v_g = 230\:V_{rms}$ and $V_o = 345\:V$)}
\end{figure*}

\begin{figure}[!htbp]
     \centering

         \includegraphics[width=0.8\linewidth]{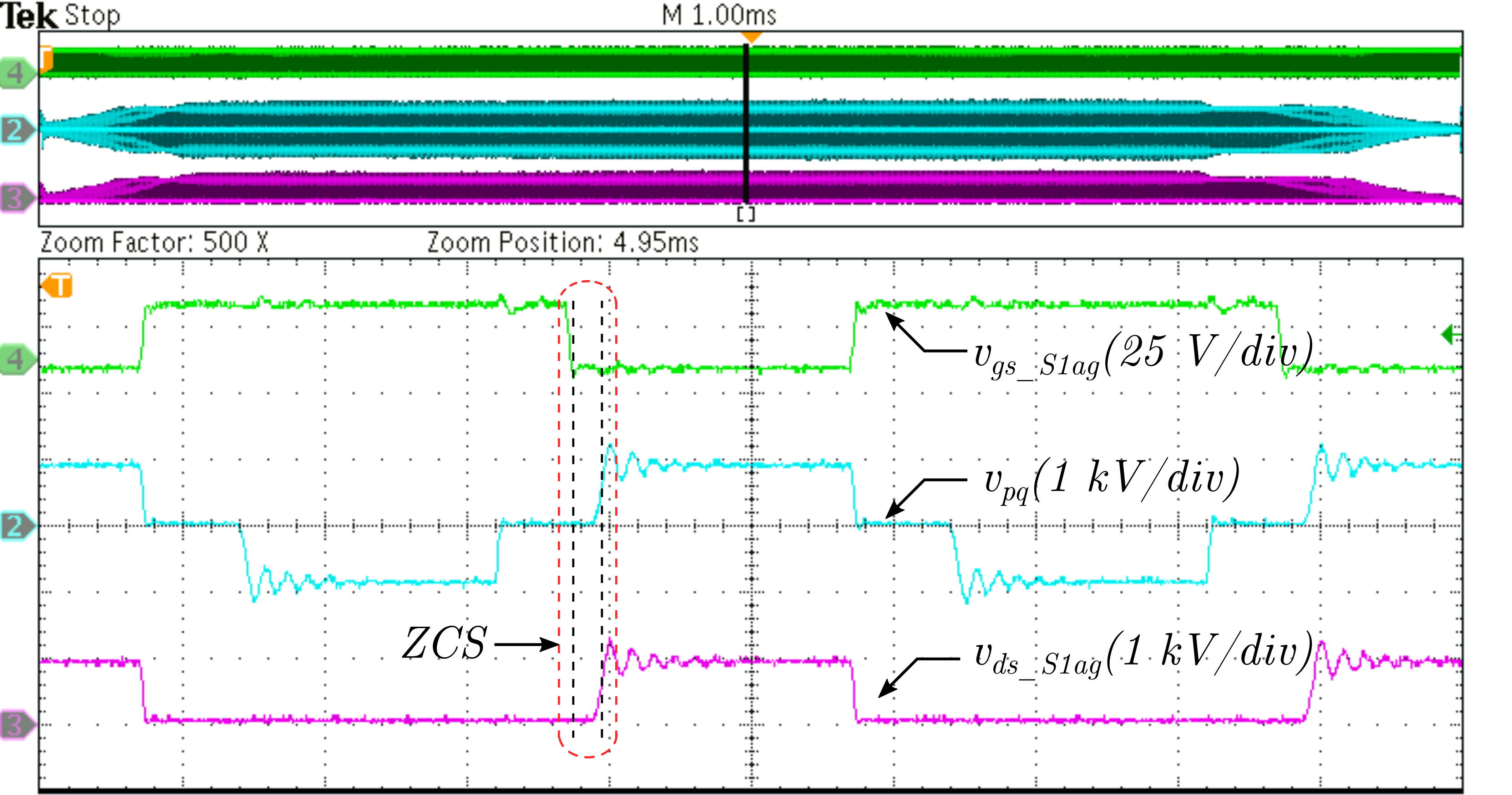}

	  \caption{Zoomed-in measured gate source voltage $v_{gs-S_{1ag}}$, drain source voltage $v_{ds-S_{1ag}}$ and the ac side HFT voltage $v_{pq}$}
	  	\label{fig:zcs}
\end{figure}

\subsection{Experimental Waveform}
Fig. \ref{fig:grid_cycle_old} shows the measured ac grid voltage $v_g$ along with the input grid current $i_g$, ac side HFT winding current $i_{lk}$ and output voltage $V_o$ for DCPSM \cite{prasannaNovelZeroCurrentSwitchingCurrentFed2013}. It can be observed that the peak value of the winding current is constant in the entire grid cycle of ac mains voltage. This is further depicted in Fig. \ref{fig:peak_zoomed_old} and \ref{fig:zero_zoomed_old} showing the zoomed waveform at the peak and near the zero crossing of the ac mains voltage respectively. The peak value of the winding current is 10.2 A. This higher circulating current near the zero crossing of ac mains voltage, attributing to higher conduction losses and deteriorating efficiency.

Fig. \ref{fig:grid_cycle_new} shows the measured ac grid voltage $v_g$, grid current $i_g$, ac side HFT winding current $i_{lk}$ and output voltage $V_o$ for proposed IDCPSM. It is clear from the figure that  converter is operating at 0.999 power factor. The total series inductor current now has a sinusoidal envelope with twice the ac mains voltage frequency. Its peak current value has reduced by almost $55\%$ from the peak Fig. \ref{fig:peak_zoomed_new} to near the zero crossing Fig. \ref{fig:zero_zoomed_new} of the ac mains voltage. 

Fig. \ref{fig:zcs} shows the zoomed waveform of gate source voltage  $v_{gs-S_{1ag}}$ and drain source voltage $v_{ds-S_{1ag}}$ of ac side switch $S_{1ag}$, along with the ac side HFT voltage $v_{pq}$. It can be observed that (dotted region) even though the gate voltage is at -5V, the switch voltage is zero, indicating that its body diode is conducting as discussed in section II. The switch naturally turns-off as the current through its body diode goes to zero achieving ZCS. 

\begin{figure}[ht]
     \centering

         \includegraphics[width=0.9\linewidth]{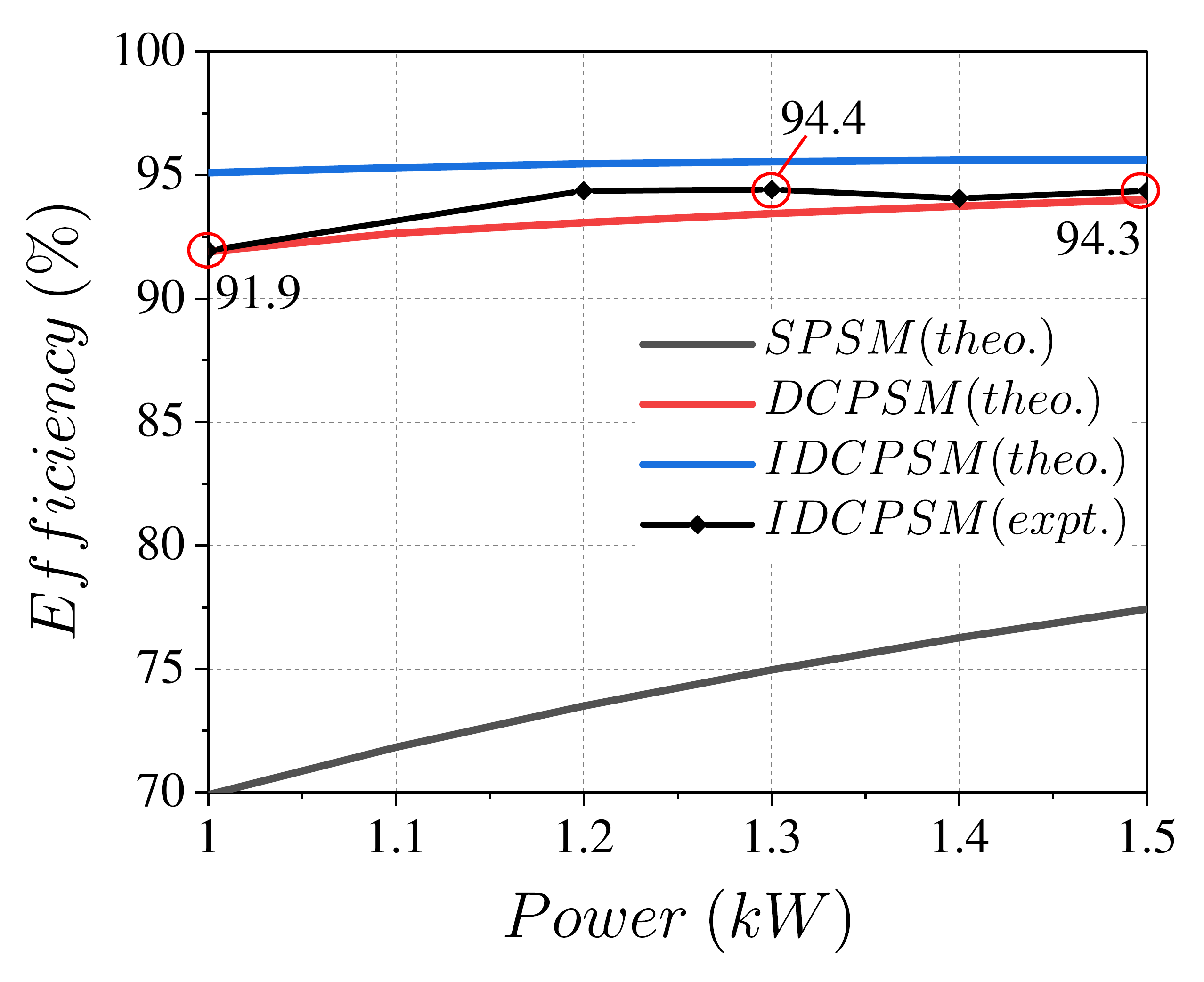}

	  \caption{Efficiency curve for different SSM at $V_o = 345\:V$ }
	  	\label{fig:efficiency}
\end{figure}

\subsection{Efficiency}

\begin{figure}[ht]
     \centering

         \includegraphics[width=0.9\linewidth]{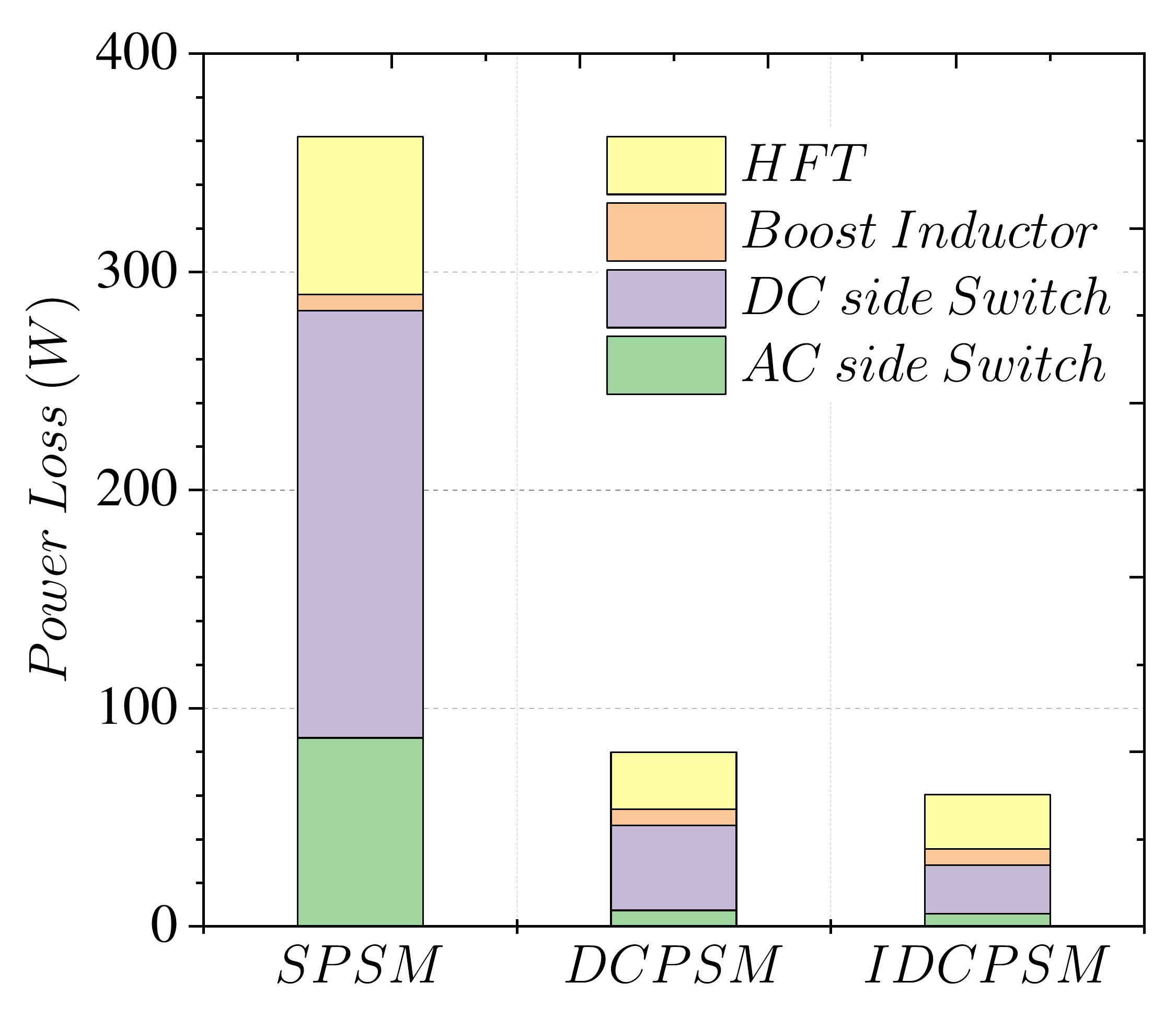}

	  \caption{Loss distribution among the components of L-L type ICFHB ac-dc converter at $P_o = 1.5\:kW $and $V_o = 345\:V$ for different SSM}
	  	\label{fig:loss_break}
\end{figure}

Fig. \ref{fig:efficiency} shows the efficiency curve for different SSM at nominal output voltage $V_o = 345\:V$. Power analyser module of Tektronix MDO3104 DSO is used for measuring power. From the figure it can be concluded that the proposed IDCPSM has the best performance with peak efficiency of $94.4 \%$. Furthermore, the efficiency curve for IDCPSM remains nearly flat across different loading conditions. An improvement of around $3 \%$ is seen at lighter loading condition $P_o = 1\:kW$ and $1.5\%$ at full load $P_o = 1.5\:kW$ from DCPSM. This improvement in performance is due to active control of the peak  circulating across the gird cycle of the ac mains voltage and at different loading conditions. The SPSM is the least efficient among the three SSM. Fig. \ref{fig:loss_break} shows the loss distributions among the components. The sum of switching and conduction loss is depicted as ac or dc side switch loss. HFT loss shows the total loss occurred in the HFT and the series inductor. The reduction in peak value of circulating current with the proposed IDCPSM reduces the conduction losses in ac side switch as depicted in the Fig. \ref{fig:loss_break}. The total loss in the dc side switch is reduced by almost $42\%$ from DCPSM. This is because both switching and conduction loss are reduced in the dc side switch. Furthermore, the total loss in the HFT is reduced by almost $20\%$ with the proposed IDCPSM as both core loss and copper loss reduces in the HFT and the series inductor.  

\section{Conclusion}
This paper proposed a IDCPSM for an L-L type ICFHB ac-dc converter. A dual control variable approach has been developed where in ac side CFHB duty is modulated to maintain UPF and dc side VFFB duty is modulated to reduce the peak circulating current in a grid cycle of ac mains voltage and also at light load conditions. ZCS turn-off of ac side switches and ZVS turn-on of dc side switches are maintained throughout the operating condition. This has significantly reduced the conduction and switching losses in both ac and dc side switches. A comparative performance analysis with different SSM reported in literature is also presented. The proposed IDCPSM provides the best performance with  peak efficiency of $94.4\%$ and almost a flat efficiency profile across different loading conditions.


%




\bibliographystyle{IEEEtran}
\bibliography{IEEEabrv,references_v1}
%
%








\end{document}